\documentclass[twocolumn]{autart}    

\usepackage{graphicx}
\usepackage{epsfig}
\usepackage{mathptmx} 
\usepackage{times} 
\usepackage{amsmath} 
\usepackage{amssymb}  
\usepackage{epstopdf} 
\usepackage{subfigure}
\usepackage{float}

\begin{document}

\begin{frontmatter}

\title{A Generic Formation Controller and State Observer for Multiple Unmanned Systems} 

\thanks[footnoteinfo]{Corresponding author R.~Dutta. Tel. +91 7619597893.}

\author[RD]{Rajdeep Dutta}\ead{rajdeepdutta.iisc@gmail.com},
\author[CJ]{Chunjiang Qian}\ead{chunjiang.qian@utsa.edu},
\author[LS]{Liang Sun}\ead{lsun@nmsu.edu},
\author[DP]{Daniel Pack}\ead{daniel-pack@utc.edu}

\address[RD]{Computational Mechanics Laboratory, Indian Institute of Science, Bangalore-560012, India}
\address[CJ]{Department of Electrical and Computer Engineering,
University of Texas at San Antonio, TX-78249, USA}
\address[LS]{Department of Mechanical and Aerospace Engineering,
New Mexico State University, NM-88003, USA}
\address[DP]{Department of Electrical Engineering, University of Tennessee at Chattanooga, Chattanooga, TN 37403-2598, USA}


\begin{abstract}                          
In this paper, we present a novel decentralized controller to drive multiple unmanned aerial vehicles (UAVs) into a symmetric formation of regular polygon shape surrounding a mobile target. The proposed controller works for time-varying information exchange topologies among agents and preserves a network connectivity while steering UAVs into a formation. The proposed nonlinear controller is highly generalized and offers flexibility in achieving the control objective due to the freedom of choosing controller parameters from a range of values. By the virtue of additional tuning parameters, i.e. fractional powers on proportional and derivative difference terms, the nonlinear controller procures a family of UAV trajectories satisfying the same control objective. An appropriate adjustment of the parameters facilitates in generating smooth UAV trajectories without causing abrupt position jumps. The convergence of the closed-loop system is analyzed and established using the Lyapunov approach. Simulation results validate the effectiveness of the proposed controller which outperforms an existing formation controller by driving a team of UAVs elegantly in a target-centric formation. We also present a nonlinear observer to estimate vehicle velocities with the availability of position coordinates and heading angles. Simulation results show that the proposed nonlinear observer results in quick convergence of the estimates to its true values.
\end{abstract}

\end{frontmatter}

\section{Introduction}
\begin{figure}[h]
 \centering
   \includegraphics[width=9.2cm,height=7.5cm]{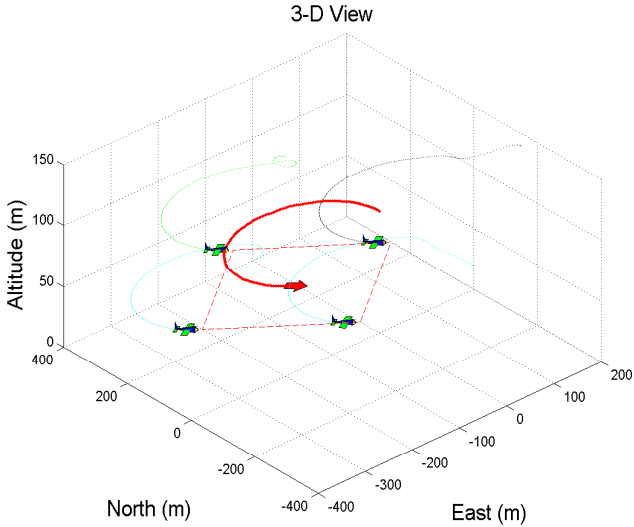}
   \caption{Symmetric square shaped target-centric formation by UAVs.}
   \label{fig:form_simulator}
\end{figure}

The study of cooperative control has become popular over the years in Unmanned Aerial Vehicle (UAV) research community, due to its variety of applications in solving real world challenging problems (\cite{Form_review}, \cite{WRen}, \cite{Raj_Springer}, \cite{Desai_form}). In this paper, we investigate a specific cooperative control task pertaining to the problem of hostile target tracking by multiple UAVs (\cite{Raj_ACC}). The formation control strategies (\cite{Raj_Springer}, \cite{RD_IROS14}) enable a team of UAVs to succeed in a target-centric symmetric formation. \cite{Raj_ACC}, \cite{Raj_Springer} proposed a control law to make a formation of UAVs around a target moving along a pre-defined trajectory, utilizing complete or incomplete target state information. It is highly challenging to design an efficient cooperative controller for a group of unmanned vehicles with non-holonomic drive \cite{nonHol_rob}, where the degrees of freedom of UAVs in movement are greater than the controllable degrees of freedom; the degrees of freedom in movement are two-dimensional position coordinates and orientations, whereas the controllable degrees of freedom are linear accelerations and turning rates.\\\\
Recently, in \cite{RD_IROS14}, we introduced a control law that drives multiple UAVs in a target-centric formation along with maintaining the dynamic connectivity of the group. The formation controllers mandate that the multi-agent network is connected at all times throughout a mission. \cite{Autm1} addressed the coordinated multi-agent formation control problem, characterized by repetition and switching network topologies, and devised a distributed iterative update algorithm based on the nearest neighbor's relative distance. In \cite{Autm2}, a distributed control law was proposed by exploiting the quantized values of the relative states between neighboring agents to solve a multi-agent consensus problem with limited inter-agent communication for static and time-varying topologies. Fig. \ref{fig:form_simulator} shows a typical scenario where four UAVs are flying in a symmetric formation around a mobile aerial target.

\noindent  The connectivity of a team of UAVs, represented as a multi-agent graph, plays a pivotal role in group dynamics, as it represents the level of the information exchange capability among all members. A network's connectivity can be determined by estimating (\cite{Fiedler_73}) the second smallest eigenvalue of the associated Laplacian matrix (\cite{Book_GraphTh}). During the motion of a multi-agent system, if the second smallest eigenvalue, known as the \textit{algebraic connectivity}, diminishes to zero then the network gets disconnected. The convergence speed of a multi-agent system dynamics also depends on the algebraic connectivity of the corresponding graph \cite{WRen}. Previous studies (\cite{Ghosh_Boyd}, \cite{Decn_control}, \cite{Pap_conn1}, \cite{Pap_conn2}) show a wide variety of research performed in maintaining and controlling the connectivity of a multi-agent graph. An efficient technique to enhance the algebraic connectivity of a graph was proposed in \cite{Ghosh_Boyd}. \cite{Pap_conn1}, \cite{Pap_conn2} developed artificial potential fields in order to maintain, increase and control connectivity of mobile robot networks. To maximize the algebraic connectivity of a proximity graph consisting of multiple agents, authors in \cite{Decn_control} developed a decentralized algorithm. However, to the best of the authors' knowledge, the problem of maintaining and controlling the connectivity of a dynamic multi-agent network pertaining to the formation control task is much more challenging and yet to be explored.\\\\
\noindent  In recent years, fractional-order nonlinear controllers (\cite{Bhat_FT1}, \cite{Bhat_FT2}) gain interests as they exhibit advantageous characteristics (\cite{FracCon_app1}) and offer more flexibility in achieving a control objective over those of conventional controllers. \cite{Bhat_FT1} had shown the finite time stability of a nonlinear fractional order differential equation. In \cite{FracCon_app1}, authors used a fractional order nonlinear controller for load frequency control to provide quality power in isolated and interconnected power plants, by exploiting controller's novel properties, such as robustness toward plant gain variations and good noise rejection capability. A brief tutorial on fractional order dynamics and control was presented in \cite{frac_tut}; this study also introduced fractional order PID controllers and their usage in industries. Another study by \cite{FracCon_app2} shows the advantages of using a fractional order PID controller over the traditional PID controller on the hydraulic turbine regulating system. \\\\
\noindent  Earlier studies (\cite{obsvCJ}, \cite{obsvDu2}) presented finite-time nonlinear observers, and established the benefit of using fractional powers in achieving fast estimates. The work of \cite{obsvCJ} considered nonlinear systems with time-varying and output-dependent coefficients, and used a recursive design method in conjunction with homogeneous domination approach for observer development. For leader-free and leader-follower based multi-agent systems, \cite{obsvDu2} designed a consensus algorithm with a foundation based on distributed finite-time observers. In \cite{compl_filter}, a fractional order complementary filter was used for orientation estimation of unmanned aerial systems with low-cost sensors, in the presence of non-Gaussian noise. Also in biomedical engineering (\cite{biomed_frac}), nonlinear fractional-order observer has fruitful applications for estimating the structure of a blood glucose-insulin with glucose rate disturbance as an unknown input.\\\\
\noindent  This work presents a generic decentralized controller for multi-UAV network to accomplish the cooperative task of surrounding a mobile target in a symmetric formation of regular polygon shape. The proposed controller has potential of generating a family of UAV trajectories with guaranteed consensus convergence. Among the family of trajectories, a specific one can be chosen according to the application requirement. The controller involves additional tuning parameters rendering more degrees of freedom in agent movements. To this end, an intelligent parameter selection scheme is adopted, which enables gradual agent movements during the multi-agent dynamics and indirectly helps in improving the dynamic network connectivity. Apart from the controller development, the present work also addresses an application of the fractional observer for quick state estimation required in formation control problems.\\\\
The rest of the paper is organized as follows. In Section \ref{sec:ProbDescrip}, we illustrate the graph theory related to multi-agent systems, formulate the problem, and state the objective of this work. Section \ref{sec:controller} presents the proposed nonlinear controller and shows the system state convergence using the Lyapunov theory. In Section \ref{sec:connectivity}, we introduce a parameter selection scheme for the nonlinear controller in order to take care of the dynamic network connectivity. In Section $5$, we describe the nonlinear state observer. Section \ref{sec:results} presents the simulation results and establishes the proposed controller's novelty in generating smooth trajectories with limited input to accomplish the formation task. Finally, Section \ref{sec:conclusion} concludes the paper.

\section{Problem Description}\label{sec:ProbDescrip}
\subsection{Multi-agent system and related Graph Theory}
A dynamic multi-agent network comprised of point-mass UAVs and a target can be analyzed mathematically with the help of a time-varying graph representation, $G(t)=(V,E(t))$, with fixed number of nodes $V$ and variable number of edges $E(t)$, where the nodes of the graph represent agents and the edges represent communication links or connections between them.

Let, ${\bf p}_i$ and ${\bf p}_j$ denote the position vectors of the $i$th and the $j$th agents, respectively, ${\bf r}_{ij}$ denotes the relative distance vector between the $i$th and $j$th agents, and $R$ is the communication range of each agent. The position vector and the velocity of the target be denoted by $\mathbf{p}_{l}=\mathbf{p}_{n+1}$ and $v_{l}=v_{n+1}$, respectively, where $(n+1)$ is the total number of agents including $n$ UAVs and one target.\\

\noindent  We assume all agents have the same communication range, and the multi-agent graph of order $(n+1)$, i.e. $G_{n+1}(t)$, is undirected with bidirectional communication links. A connection between a pair of agents $(i,j)$ exists when the relative distance between them is within the communication range. In other words, an agent $j$ becomes agent $i$'s neighbor, where the neighborhood of $i$th agent is defined as
\begin{align}\label{eqn:nghbor}
& N_i(t) = \{ j\in V,j\neq i, j\sim i: \| \mathbf{r}_{ij} \|=\parallel \mathbf{p}_i -\mathbf{p}_j \parallel\leq R \}.
\end{align}

\noindent  The time-varying connections among multiple agents are captured using a time-varying Laplacian matrix \cite{Book_GraphTh}. The Laplacian elements depend on the relative distances between the corresponding pair of nodes. A state dependent Laplacian matrix of order $(n+1)$, is given by
\begin{equation}\label{eqn:DminusA}
L_{n+1}(\textbf{s}(t)) = D_{n+1}(\textbf{s}(t)) - A_{n+1}(\textbf{s}(t))~,
\end{equation}
where $\textbf{s}(t)$ denotes the vector consisting of agent states, $D_{n+1}$ is the Degree matrix, and $A_{n+1}$ is the Adjacency matrix.

In accordance with an exponential communication model \cite{Decn_control}, the elements of a weighted Adjacency matrix ($a_{ij}$) are given as
\begin{eqnarray}\label{eqn:adj}
a_{ij}(t) =
\left\{
\begin{array}{lll}
& e^{- \frac{\sigma}{R} r_{ij} } ~~~\mbox{if}~~i \neq j ~~\& ~~ r_{ij}  \leq R ~ ; \\
& 0 ~~~\mbox{if}~~i \neq j ~~\& ~~ r_{ij}  \geq R ~ ; \\
& 0 ~~~\mbox{if}~~~ i=j ~ ,
\end{array}
\right.
\end{eqnarray}
where $r_{ij} = \| \mathbf{r}_{ij} \|$ denotes the relative distance magnitude between $(i,j)$, and $\sigma$ is a large positive constant representing the decay rate of the communication quality over distance. The constant $\sigma$ in Equation (\ref{eqn:adj}) is selected in such a way that it ensures $j \in N_i$ only if $ r_{ij} \leq R$, which in turn means that the connection is lost beyond an upper bound on the range $R$, and it varies exponentially within this range.

\noindent The Degree matrix is a diagonal matrix with elements given by
\begin{equation*}
d_i=\sum \limits_{j(\neq i)=1}^{n+1} a_{ij}~.
\end{equation*}

\noindent Using Equation (\ref{eqn:DminusA}), the Laplacian matrix ($L_{n+1}$) elements are expressed as
\begin{equation}\label{eqn:lapc}
l_{ij}(t)=
\left\{
\begin{array}{lll}
-a_{ij} ~\mbox{for}~i\neq j; \\
\sum \limits_{j(\neq i)=1}^{n+1} a_{ij} ~\mbox{for}~ i=j.
\end{array}
\right.
\end{equation}
The Laplacian, $L(t)$, for an undirected graph, $G(t)$, is a positive semi-definite and symmetric matrix. The smallest eigenvalue of the Laplacian matrix, $\lambda_1$, is zero, and the corresponding eigenvector is $\mathbf{1}$. The second smallest eigenvalue of $L(t)$, i.e. $\lambda_2(t)$, is called the \textit{algebraic connectivity} of the graph, and the corresponding eigenvector is known as the Fiedler vector \cite{Fiedler_73}. This eigenvalue, $\lambda_2(t)$, provides a quantitative measure of the entire network connectivity and the level of information sharing capability among agents at time $t$; $\lambda_2=0$ indicates that the graph has no spanning tree and so the information flow path among agents has discontinuity, whereas $\lambda_2 > 0$ indicates a connected graph with at least one spanning tree, and the amount of connections (number of spanning trees) increases as $\lambda_2$ increases.

In a multi-UAV graph, the mapping from the agent topological configuration space to the connectivity is an example of multiple-to-one or surjective mapping, which in turn means that the problem of identifying a particular topology from just the connectivity measure is not unique and there can exist distinct multi-agent topologies of identical connectivity \cite{toplg_isoCon}.

\subsection{Problem formulation}
\noindent  Consider a multi-agent network of order $(n+1)$ consisting of $n$ UAVs and one target. Every UAV shares information about its states and control inputs with its neighbor(s). The target follows a predefined trajectory and does not utilize information gathered from UAVs, whereas UAVs utilize the complete target information. All agents are assumed to fly at a fixed altitude.

\noindent  Assuming a network of nonholonomic point-agents engaged in tracking a mobile target, the dynamics of the $i$th UAV is governed by the following equations.
\begin{align}\label{eqn:model}
& \dot{{\bf p}_i}=\left[
           \begin{array}{c}
             \dot{x_i} \\
             \dot{y_i}
           \end{array}
          \right]=
          \left[
           \begin{array}{c}
             v_i \cos \phi_i \\
             v_i \sin \phi_i
           \end{array}
          \right] \\
&  \ddot{{\bf p}}_i=\left[
             \begin{array}{cc}
               \cos \phi_i &  -v_i \sin \phi_i \\
               \sin \phi_i &  ~~v_i \cos \phi_i \\
             \end{array}
           \right]
           \left[
             \begin{array}{c}
               \dot{v_i} \\
               \dot{\phi_i} \\
             \end{array}
           \right] \\
& \ddot{\mathbf{p}}_i=M_i \mathbf{u}_i~,~~ \\
&      {\bf u}_i \triangleq \left[
           \begin{array}{c}
             \dot{v_i} \\
             \dot{\phi_i}
           \end{array}
          \right],~~   M_i \triangleq \left[
             \begin{array}{cc}
               \cos \phi_i &  -v_i \sin \phi_i \\
               \sin \phi_i &  ~~v_i \cos \phi_i \\
             \end{array}
           \right]~, \nonumber
\end{align}
where  $\mathbf{p}_i=[x_i,~y_i]^T$, $v_i$ , $\phi_i$ , and $\mathbf{u}_i$ denote position, velocity, heading angle, and control input of the $i$th UAV, respectively. The full state vector of $i$th agent can be represented as: $\mathbf{s}_i(t)=[\mathbf{p}_i^T, v_i, \phi_i]^T$.  The objective of a symmetric formation control problem is to drive all UAVs at a constant distance $\delta$ and a relative angle $\psi_i$ with respect to the target.

\begin{figure}[h]
 \centering
   \includegraphics[width=5.7cm,height=6.5cm]{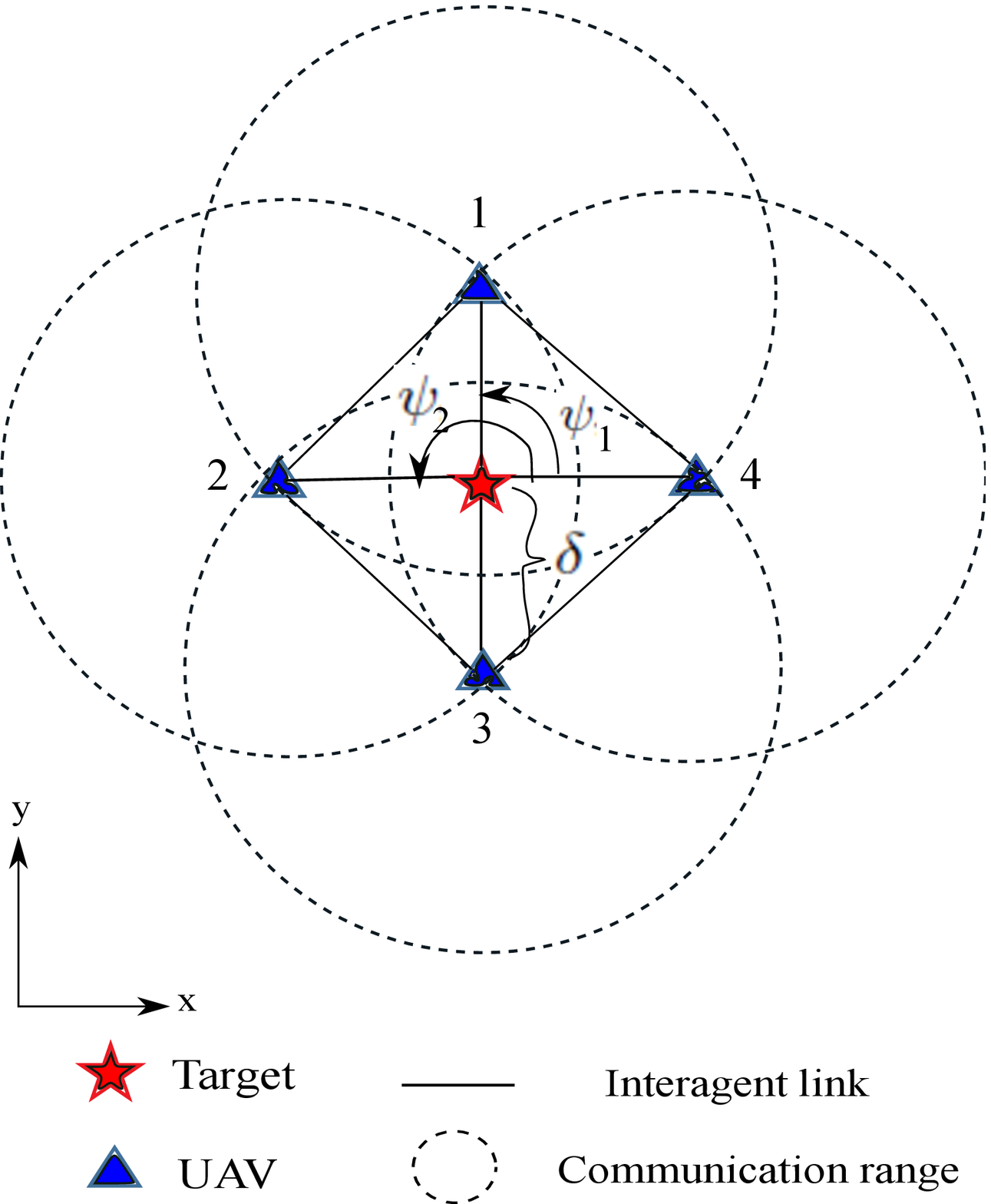}
   \caption{A scenario: Desired formation for four UAVs around one target. }
   \label{fig:formDes}
\end{figure}

\noindent  The aim of the present work is to achieve a target-centric symmetric formation of UAVs with capabilities to adjust the connectivity. The control objective can be expressed mathematically as follows.
\begin{align}\label{eqn:obj}
& {\bf p}_i(t)-{\bf p}_t(t) \rightarrow  {\bf P}_i \\
& \dot{{\bf p}}_i(t) - \dot{{\bf p}}_t(t) \rightarrow 0  \\
& \psi_{i+1} - \psi_{i}= \frac{2\pi}{n}~,
\end{align}
where ${\bf P}_i= \delta [\cos \psi_i ~~ \sin \psi_i]^T$ is the desired location vector of the $i$th agent for a pre-specified $\delta$. In Fig. \ref{fig:formDes}, we show a desired target-centric symmetric formation of four UAVs.
It is noteworthy that the desired final separations between UAVs and target are consistent, indicating that the desired formation is a target-centric symmetric formation of regular polygon shape.

\section{Decentralized Formation Controller}\label{sec:controller}
The target-centric formation controller utilizing complete target information \cite{Raj_Springer} is explained in the following. In order to drive the relative distance state ($\mathbf{p}_i(t) - \mathbf{p}_j(t)$) to the desired separation ($\mathbf{P}_i - \mathbf{P}_j$), the $i$th agent uses a control input $\mathbf{u}_i$. The controller equation considering only two agents $i$ and $j$ is given by
\begin{equation}\label{eqn:ConIPij}
 M_i \mathbf{u}_i =  \ddot{\mathbf{p}}_j - k_1 (\hat{\mathbf{p}}_i-\hat{\mathbf{p}}_j) - k_2 (\dot{\mathbf{p}}_i-\dot{\mathbf{p}}_j),~~i\neq j~,
\end{equation}
where $k_1$ and $k_2$ are positive controller gains, and $\hat{\mathbf{p}}_i= \mathbf{p}_i - \mathbf{P}_i$. Considering the multi-agent network consensus \cite{WRen}, the control input $\mathbf{u}_i$ for the $i$th UAV $i=1,2,...,n$ is given by
\begin{align}\label{eqn:controlIP}
& \mathbf{u}_i= \frac{M_i^{-1}}{\sum \limits_{j \in N_i} a_{ij}} \sum \limits_{j \in N_i} a_{ij} [\ddot{\mathbf{p}}_j - k_1 (\hat{\mathbf{p}}_i-\hat{\mathbf{p}}_j) - k_2 (\dot{\mathbf{p}}_i-\dot{\mathbf{p}}_j)].
\end{align}
Note that the control input (\ref{eqn:controlIP}) contains a singularity when the $i$th UAV velocity $v_i$ goes to zero; however, this condition never arises due to the lower bound on the velocity. The present work considers time-varying topologies among agents along with dynamic controller weights, $a_{ij}(t)$, as shown in Equation (\ref{eqn:adj}).
In this work, we modify the inter-agent dynamics shown in Equation(\ref{eqn:ConIPij}) by introducing a fractional order nonlinear finite time stable controller \cite{Bhat_FT1}, as shown in the following.
\begin{align}\label{eqn:ConIPij_new}
M_i \mathbf{u}_i = \ddot{\mathbf{p}}_j - k_1 (\hat{\mathbf{p}}_i-\hat{\mathbf{p}}_j)^{\alpha_1} - k_2 (\dot{\mathbf{p}}_i-\dot{\mathbf{p}}_j)^{\alpha_2},~~i\neq j
\end{align}
where $\alpha_1=(2\tau +1)$, and $\alpha_2=\frac{(2\tau +1)}{(\tau+1)}$ with $-\frac{1}{2} < \tau \leq 0$ are fractional powers to the proportional and derivative difference terms, respectively. The proposed inter-agent control law shown in Equation (\ref{eqn:ConIPij_new}), becomes the linear case shown in Equation (\ref{eqn:ConIPij}) when $\tau=0$. After incorporating the consensus protocol for controlling group dynamics, the controller for multi-agent system is obtained as follows.
\begin{align}\label{eqn:controllerCJ}
& \mathbf{u}_i= \frac{M_i^{-1}}{\sum \limits_{j \in N_i} a_{ij}(t)} \sum \limits_{j \in N_i} a_{ij}(t) [\ddot{\mathbf{p}}_j - k_1 (\hat{\mathbf{p}}_i-\hat{\mathbf{p}}_j)^{\alpha_1} - k_2 (\dot{\mathbf{p}}_i-\dot{\mathbf{p}}_j)^{\alpha_2}]
\end{align}
The decentralized controller proposed in Equation (\ref{eqn:controllerCJ}) offers advantageous tuning protocols, and a proper adjustment of the controller parameters, including gains ($k_1,~k_2$) and powers ($\alpha_1,~\alpha_2$), facilitates in generating desired UAV trajectories.

\noindent {\bf Theorem 1.} If every agent in a team of UAVs is governed by the dynamics in Equations (5)-(7) with the control input $\mathbf{u}_i$ in Equation (\ref{eqn:controllerCJ}), then $ (\hat{\mathbf{p}}_i-\hat{\mathbf{p}}_t) \rightarrow \mathbf{P}_i ~\text{and}~ \dot{\mathbf{p}}_i \rightarrow \dot{\mathbf{p}}_t, ~\forall~i=1,2,...,n ~\text{as}~ t \rightarrow \infty$, if and only if the multi-agent graph $G_{n+1}$ has at least one spanning tree at all times.\\

\noindent {\bf Proof.}  We first show that the double integrator system dynamics between any pair of agents, $\dot{\mathbf{X}}_1 = \mathbf{X}_2,~ \dot{\mathbf{X}}_2 = \mathbf{U}$ for $\mathbf{X}_1 = (\hat{\mathbf{p}}_i-\hat{\mathbf{p}}_j),~\text{where}~\mathbf{X}_1, \mathbf{X}_2, \mathbf{U} \in \mathbb{R}^2$, is globally asymptotically stable under the fractional order control input $\mathbf{U} = - k_1  \text{sign}(\mathbf{X}_1)|\mathbf{X}_1|^{\alpha_1} - k_2  \text{sign}(\mathbf{X}_2)|\mathbf{X}_2|^{\alpha_2}$, where $\alpha_1=(2\tau +1)~\text{and}~\alpha_2=\frac{(2\tau +1)}{(\tau+1)}$ with $ -\frac{1}{2} < \tau \leq 0$. Under control law $\mathbf{U}$, the relative state $\mathbf{X}_1$ dynamics follows a nonlinear $2$nd order differential equation given as
\begin{align}\label{eqn:nonlnDE}
\ddot{\mathbf{X}}_1  + k_2  \text{sign}(\dot{\mathbf{X}}_1)|\dot{\mathbf{X}}_1|^{\alpha_2} + k_1  \text{sign}(\mathbf{X}_1)|\mathbf{X}_1|^{\alpha_1} = \mathbf{0}.
\end{align}
\noindent  The inter-agent relative distance dynamics, between any pair of neighboring agents $i$ and $j$ under control law $\mathbf{U}$, is governed by the following state space equations.
\begin{align}\label{eqn:ModelStSp}
\left\{
\begin{array}{l l}
    \dot{\mathbf{X}}_1 = \mathbf{X}_2 \\
    \dot{\mathbf{X}}_2 = - k_1  \text{sign}(\mathbf{X}_1)|\mathbf{X}_1|^{\alpha_1} - k_2  \text{sign}(\mathbf{X}_2)|\mathbf{X}_2|^{\alpha_2}.
\end{array}\right.
\end{align}
In order to show the global asymptotic stability of the system in Equation (\ref{eqn:ModelStSp}), we choose a Lyapunov candidate function without loss of generality. The positive definite candidate function ($V>0 ~\forall~ \mathbf{X}_1,\mathbf{X}_2 \neq \mathbf{0}$), is as follows
\begin{align}\label{eqn:candLypv}
& V(\mathbf{X}_1,\mathbf{X}_2)= \frac{1}{2} \mathbf{Q}^T \mathbf{Q} \\ \label{eqn:candP}
& \text{where,}~~ \mathbf{Q} = \left( \frac{1}{2}\right)  \mathbf{X}_2^2 + \left( \frac{k_1}{\alpha_1+1}\right) |\mathbf{X}_1|^{\alpha_1+1}~.
\end{align}
The positive vector $\mathbf{Q}$ in Equation (\ref{eqn:candP}) involves element-wise vector product, i.e. Schur product or Hadamard product (\cite{Book_HornJohn}) denoted by $\circ$, in powers of $\mathbf{X}_1$ and $\mathbf{X}_2$. Taking time derivatives on both sides of Equation (\ref{eqn:candP}), we obtain
\begin{align}\label{eqn:candLypDerv}
 \dot{\mathbf{Q}}= \mathbf{X}_2 \circ \dot{\mathbf{X}}_2 + k_1 \text{sign}(\mathbf{X}_1)|\mathbf{X}_1|^{\alpha_1} \circ \dot{\mathbf{X}}_1~.
\end{align}
Now, we substitute Equations (\ref{eqn:ModelStSp}) and (\ref{eqn:candP}) into Equation (\ref{eqn:candLypDerv}) to determine $\dot{\mathbf{Q}}$ as
\begin{align*}
& \dot{\mathbf{Q}}= \mathbf{X}_2 \circ (- k_1  \text{sign}(\mathbf{X}_1)|\mathbf{X}_1|^{\alpha_1} - k_2  \text{sign}(\mathbf{X}_2)|\mathbf{X}_2|^{\alpha_2})   \\
& +  k_1 \text{sign}(\mathbf{X}_1)|\mathbf{X}_1|^{\alpha_1} \circ \mathbf{X}_2 \\
& \Rightarrow \dot{\mathbf{Q}}= -k_2 |\mathbf{X}_2|^{\alpha_2+1}< 0 ~\forall~ \mathbf{X}_2 \neq \mathbf{0}~.
\end{align*}
Taking time-derivatives on both sides of Equation (\ref{eqn:candLypv}), we obtain
\begin{align}\label{eqn:LypCandDerv}
& \dot{V}(\mathbf{X}_1,\mathbf{X}_2) = \mathbf{Q}^T \dot{\mathbf{Q}} = -k_2 \mathbf{Q}^T (|\mathbf{X}_2|^{\alpha_2+1})~.
\end{align}
In Equation (\ref{eqn:LypCandDerv}), the vector $\mathbf{Q}$ being positive, makes the time derivative of the Lyapunov candidate $\dot{V}$ negative, which implies the asymptotical stability of the system under control input $\mathbf{U}$, which in turn means $\mathbf{p}_i - \mathbf{p}_j \rightarrow \mathbf{P}_i - \mathbf{P}_j~\text{and}~ \dot{\mathbf{p}}_i \rightarrow \dot{\mathbf{p}}_j$ as $ t \rightarrow \infty$.

Now, we define the relative position error vector for all UAVs as $\mathbf{e}_p=\hat{\mathbf{p}}_{f}-(\mathbf{1} \otimes \mathbf{p}_{l}) \in \Re^{2n \times 1},~\text{where}~ \hat{\mathbf{p}}_{f}(t)=\mathbf{p}_{f}(t)-\mathbf{P}_{f} \in \Re^{2n \times 1}$; $\mathbf{p}_{f}(t)= [\mathbf{p}_1^T,\mathbf{p}_2^T,....,\mathbf{p}_n^T]^T \in \Re^{2n \times 1}$ contains all UAV position states, $\mathbf{P}_{f}= [\mathbf{P}_1^T,\mathbf{P}_2^T,....,\mathbf{P}_n^T]^T \in \Re^{2n\times 1}$ contains all UAV desired locations. We denote $\mathbf{p}_{ft}=[\mathbf{p}_f^T,\mathbf{p}_t^T]^T \in \Re^{2(n+1) \times 1}$ as the full position state vector, and $\mathbf{p}_{ft}=[\mathbf{P}_f^T,\mathbf{p}_t^T]^T \in \Re^{2(n+1) \times 1}$ as the full desired location vector for all agents.

\noindent  The dynamics of the $i$th agent governed by the proposed controller in Equation (\ref{eqn:ConIPij_new}) can be expressed as follows:
\begin{align}\label{eqn:dynmEnhanced}
 \sum\limits_{j \in N_i} a_{ij} (\ddot{\mathbf{p}}_i - \ddot{\mathbf{p}}_j) = - k_2 \sum_{j} a_{ij} (\dot{\mathbf{p}}_i - \dot{\mathbf{p}}_j)^{\alpha_2} - k_1 \sum_{j} a_{ij}(\hat{\mathbf{p}}_i - \hat{\mathbf{p}}_j)^{\alpha_1}.
\end{align}
Equation (\ref{eqn:dynmEnhanced}) can further be simplified by manipulating with the powers $\alpha_1$ and $\alpha_2$ on the difference terms as
\begin{align}\label{eqn:dynmMASineq}\nonumber
& \sum_j a_{ij} (\ddot{\mathbf{p}}_i - \ddot{\mathbf{p}}_j) + k_2 \sum_j a_{ij} (\dot{\mathbf{p}}_i^{\alpha_2} - \dot{\mathbf{p}}_j^{\alpha_2}) + k_1 \sum_j a_{ij}(\hat{\mathbf{p}}_i^{\alpha_1} - \hat{\mathbf{p}}_j^{\alpha_1}) \\
& + k_2 \mathbf{\delta}_{2i} + k_1 \mathbf{\delta}_{1i} = \mathbf{0}~,
\end{align}
where $\mathbf{\delta}_{1i}= \sum\limits_j a_{ij} \{ (\hat{\mathbf{p}}_i - \hat{\mathbf{p}}_j)^{\alpha_1} -  (\hat{\mathbf{p}}_i^{\alpha_1} - \hat{\mathbf{p}}_j^{\alpha_1}) \}$, and $ \mathbf{\delta}_{2i} = \sum\limits_j a_{ij} \{ (\dot{\mathbf{p}}_i - \dot{\mathbf{p}}_j)^{\alpha_2} -  (\dot{\mathbf{p}}_i^{\alpha_2} - \dot{\mathbf{p}}_j^{\alpha_2}) \}$ are two residual terms in $\Re^{2 \times 1}$.

\noindent  For all agents present in the network, the Equation (\ref{eqn:dynmMASineq}) is converted into a matrix form by using the Laplacian matrix representing the multi-agent time varying communication topology.
\begin{align}\label{eqn:allMASMatrix}\nonumber
& (L_{n+1} \otimes I_2)\ddot{\mathbf{p}}_{ft} + k_2  (L_{n+1} \otimes I_2)\dot{\mathbf{p}}_{ft}^{\alpha_2} + k_1 (L_{n+1} \otimes I_2)\hat{\mathbf{p}}_{ft}^{\alpha_1}\\
& + k_2 \mathbf{\mu}_2 + k_1 \mathbf{\mu}_1 = \mathbf{0}~,
\end{align}
where $\mathbf{\mu}_1 = [\mathbf{\delta}_{11}^T,...,\mathbf{\delta}_{1 l}^T]^T \in \Re^{2(n+1) \times 1}$ and $\mathbf{\mu}_2 = [\mathbf{\delta}_{21}^T,...,\mathbf{\delta}_{2 l}^T]^T \in \Re^{2(n+1) \times 1}$. After segregating the target-agent connection part (last row and column) from the network of $(n+1)$ agents, the Laplacian matrix of order $(n+1)$ can be reconstructed as
\begin{align}\label{eqn:LaplSep}
L_{n+1} = \left[ \begin{array}{cc}
L_n+B_n & -\mathbf{b}_n \\
-\mathbf{b}_n^T &  \sum\limits_{i=1}^{n} a_{i l}
\end{array} \right] \in \Re^{(n+1)\times(n+1)}~,
\end{align}
where $L_n \in \Re^{n \times n}$ is the Laplacian matrix representing communication topology among $n$ agents, $\mathbf{b}_n=[a_{12},a_{13},...,a_{1n}]^T \in \Re ^{n \times 1}$ and $B_n=\text{diag}(\mathbf{b}) \in \Re^{n \times n}$.

\noindent For clarity of understanding, we now explain one of the terms in Equation (\ref{eqn:allMASMatrix}), as follows.
\begin{align*}
& (L_{n+1} \otimes I_2)\hat{\mathbf{p}}_{ft}^{\alpha_1}= \left[ \begin{array}{cc}
L_n+B_n & -\mathbf{b}_n \\
-\mathbf{b}_n^T &  \sum\limits_{i=1}^{n} a_{i l}
\end{array} \right] \left[
                      \begin{array}{c}
                        \hat{\mathbf{p}}_{f}^{\alpha_1} \\
                         \hat{\mathbf{p}}_{t}^{\alpha_1} \\
                      \end{array}
                    \right] \\
& =                  \left[
                      \begin{array}{c}
                        ((L_n+B_n) \otimes I_2) (\hat{\mathbf{p}}_f^{\alpha_1} - (\mathbf{1} \otimes \hat{\mathbf{p}}_t^{\alpha_1}))  \\
                          \sum\limits_{i=1}^{n} a_{i l} (\hat{\mathbf{p}}_t^{\alpha_1} - \hat{\mathbf{p}}_i^{\alpha_1})\\
                      \end{array}
                      \right] \because ~ (L_n+B_n)\mathbf{1}= \mathbf{b}_n.
\end{align*}
In this work, our concern is to design controllers for all UAVs $i=1,2,...,n$, so we exclude the last row from Equation (\ref{eqn:allMASMatrix}), and by applying the above Laplacian matrix construction in Equation (\ref{eqn:LaplSep}), we obtain
\begin{align}\label{eqn:fullErrDEq}\nonumber
& [ (L_n+B_n) \otimes I_2] \{ (\ddot{\mathbf{p}}_f - (\mathbf{1} \otimes \ddot{\mathbf{p}}_t)) + k_2 (\dot{\mathbf{p}}_f^{\alpha_2} - (\mathbf{1} \otimes \dot{\mathbf{p}}_t^{\alpha_2}))+ \\
& k_1(\hat{\mathbf{p}}_f^{\alpha_1} - (\mathbf{1} \otimes \hat{\mathbf{p}}_t^{\alpha_1})) \} + k_2 \mathbf{\mu}_2 |_{1:2n} + k_1 \mathbf{\mu}_1 |_{1:2n} = \mathbf{0} .
\end{align}
By using the definition of multi-agent relative state error $\mathbf{e}_p$, Equation (\ref{eqn:fullErrDEq}) can be rewritten as
\begin{align}\label{eqn:errDynmPre}
& [ (L_n+B_n) \otimes I_2]  (\ddot{\mathbf{e}}_p + k_2 \dot{\mathbf{e}}_p^{\alpha_2} + k_1 \mathbf{e}_p^{\alpha_1})  +  k_2 \mathbf{\xi}_2 +  k_1 \mathbf{\xi}_1 =\mathbf{0}~,
\end{align}
where two residual terms ${\bf \xi}_1(t),~ {\bf \xi}_2(t) \in \Re^{2n \times 1}$ are defined as
\begin{align*}
& {\bf \xi}_1 - \mathbf{\mu}_1^{'}= L_{2n}^{'}~.~ \{(\hat{\mathbf{p}}_f^{\alpha_1} - \mathbf{1} \otimes \hat{\mathbf{p}}_t^{\alpha_1})-(\hat{\mathbf{p}}_f - \mathbf{1} \otimes \hat{\mathbf{p}}_t)^{\alpha_1}\} \\
& \text{or}~~ {\bf \xi}_1 |_i = \sum\limits_j \{ (\hat{\mathbf{p}}_i - \hat{\mathbf{p}}_j)^{\alpha_1} -(\hat{\mathbf{p}}_i - \hat{\mathbf{p}}_t)^{\alpha_1} +(\hat{\mathbf{p}}_j - \hat{\mathbf{p}}_t)^{\alpha_1} \}, ~\text{and}\\
& {\bf \xi}_2 - \mathbf{\mu}_2^{'} = L_{2n}^{'}~.~ \{(\dot{\mathbf{p}}_f^{\alpha_1} - \mathbf{1} \otimes \dot{\mathbf{p}}_t^{\alpha_1})-(\dot{\mathbf{p}}_f - \mathbf{1} \otimes \dot{\mathbf{p}}_t)^{\alpha_1}\} \\
& \text{or}~~ {\bf \xi}_2 |_i = \sum\limits_j \{ (\dot{\mathbf{p}}_i - \dot{\mathbf{p}}_j)^{\alpha_1} -(\dot{\mathbf{p}}_i - \dot{\mathbf{p}}_t)^{\alpha_1} +(\dot{\mathbf{p}}_j - \dot{\mathbf{p}}_t)^{\alpha_1} \},
\end{align*}
where $\mathbf{\mu}_1^{'}=\mathbf{\mu}_1 |_{1:2n}$, $\mathbf{\mu}_2^{'}=\mathbf{\mu}_2 |_{1:2n}$, and $L_{2n}^{'}=[(L_n+B_n)\otimes I_2]$. The matrix $(L_n+B_n)$ is invertible with the assumption that the network has at least one spanning tree or the network is connected at all times. Thus, Equation (\ref{eqn:errDynmPre}) leads to the multi-agent error dynamics as
\begin{align}\label{eqn:errDynmFin}
& (\ddot{\mathbf{e}}_p + k_2 \dot{\mathbf{e}}_p^{\alpha_2} + k_1 \mathbf{e}_p^{\alpha_1}) = - [(L_n+B_n)\otimes I_2]^{-1}  (k_1 \mathbf{\xi}_1 + k_2 \mathbf{\xi}_2) .
\end{align}
It is significant to remember the fact that the spanning tree condition on a multi-agent graph allows the existence of error dynamics (\ref{eqn:errDynmFin}).

Residual terms $\mathbf{\xi}_1$ and $\mathbf{\xi}_2$ in Equation (\ref{eqn:errDynmFin}) tend to zero as the powers $\alpha_1$ and $\alpha_2$ tend to one, and $\mathbf{\xi}_1=\mathbf{\xi}_2=0$ when $\alpha_1=\alpha_2=1$ for $\tau=0$. In case $\mathbf{\xi}_1,\mathbf{\xi}_2 \approx 0$, then the nonlinear error dynamics becomes
\begin{align}\label{eqn:errDynmFinApprox}
& \ddot{\mathbf{e}}_p + k_2 \dot{\mathbf{e}}_p^{\alpha_2} + k_1 \mathbf{e}_p^{\alpha_1} \approx 0 .
\end{align}
According to the theory of finite-time stability of homogeneous systems (\cite{Bhat_FT2}), the convergence of the approximated error dynamics in Equation (\ref{eqn:errDynmFinApprox}) can be shown using a continuously differentiable Lyapunov candidate $V$ that satisfies the following inequality:
\begin{align}\label{ineq:Lypv}
(\dot{V}+c V^{\beta}) \leq 0 ~; ~ ~\forall~ c,\beta \in \Re,~c>0,~ \beta \in (0,~1).
\end{align}
In the error dynamics of Equation (\ref{eqn:errDynmFin}), the presence of the residuals on the right hand side of Equation (\ref{eqn:errDynmFin}) is a major concern regarding the closed-loop system convergence. The stability is guaranteed if the residual terms do not have substantial effects in the error dynamics.

The nonlinear error dynamics in Equation (\ref{eqn:errDynmFin}) can be represented in state space form as follows.
\begin{align}\label{eqn:NLerrSP}
\left\{
\begin{array}{l l}
    \dot{\mathbf{z}}_1 = \mathbf{z}_2 \\
    \dot{\mathbf{z}}_2 = - k_1  \text{sign}(\mathbf{z}_1)|\mathbf{z}_1|^{\alpha_1} - k_2  \text{sign}(\mathbf{z}_2)|\mathbf{z}_2|^{\alpha_2} + \mathbf{r}_s~.
\end{array}\right.
\end{align}
In Equation (\ref{eqn:NLerrSP}), $\mathbf{z}_1=\mathbf{e}_p$, $\mathbf{z}_2=\dot{\mathbf{e}}_p$, and $\mathbf{r}_s= -[(L_n+B_n)\otimes I_2]^{-1}  (k_1 \mathbf{\xi}_1 + k_2 \mathbf{\xi}_2)$. The error state vector is $\mathbf{Z}=[\mathbf{z}_1^T,~\mathbf{z}_2^T]^T$. The nonlinear error dynamics in state space form in Equation (\ref{eqn:NLerrSP}) can be further simplified as
\begin{align}\label{eqn:NLerrSP1}
 \dot{\mathbf{Z}}=\mathbf{F}(\mathbf{Z})+\mathbf{R}(\mathbf{Z}),
\end{align}
where $\mathbf{F}=[\mathbf{z}_2^T~,~ - k_1  \text{sign}(\mathbf{z}_1)|\mathbf{z}_1^T|^{\alpha_1} - k_2  \text{sign}(\mathbf{z}_2)|\mathbf{z}_2^T|^{\alpha_2}]^T$ is the homogeneous part of the nonlinear error dynamics, and $\mathbf{R}=[\mathbf{0}^T,~\mathbf{r}_s^T]^T$ is the residual part. Now, on the basis of existence (\cite{CJ_pap1}) of a continuously differentiable Lyapunov candidate $V_1$,  the time derivative of Lyapunov function along the nonlinear dynamics shown in Equation (\ref{eqn:NLerrSP1}), $\dot{V}_1$, satisfies the following equation:
\begin{align}\label{eqn:dotV1}
\dot{V}_1 = \frac{\partial V_1}{ \partial \mathbf{Z}} \mathbf{F} + \frac{\partial V_1}{ \partial \mathbf{z}_2}   \mathbf{r}_s~.
\end{align}
The first term on the right hand side of Equation (\ref{eqn:dotV1}) is asymptotically stable \cite{CJ_pap1}, as $\frac{\partial V_1}{ \partial \mathbf{Z}} \mathbf{F} \leq - c_1 V_1^{\gamma}$ for $c_1>0$ and $\gamma \in (0,1)$; however, the second term on the right hand side of Equation (\ref{eqn:dotV1}) is our main concern towards showing the stability of the entire nonlinear error dynamics shown in Equation (\ref{eqn:NLerrSP1}). The time derivative of the Lyapunov candidate, $\dot{V}_1$, follows an inequality shown below:
\begin{align}\label{eqn:ineqV1dot}
  \dot{V}_1 \leq  - c_1 V_1^{\gamma} + \frac{\partial V_1}{\partial \dot{\mathbf{e}}_p} \mathbf{r}_s~.
\end{align}
From this point, the following intends to show that the effect of the residual term, $\mathbf{r}_s$, is negligible. With the help of matrix algebra (\cite{Book_HornJohn}), we attempt to examine the effect of $\mathbf{r}_s$ in the following. By using the definition of $\mathbf{r}_s$ and the properties of vector norm, we obtain 
\begin{align}\label{eqn:normRs}
  \| \mathbf{r}_s \| \leq  \|[(L_n+B_n)\otimes I_2]^{-1}\|  \| k_1 \mathbf{\xi}_1 + k_2 \mathbf{\xi}_2 \|~.
\end{align}
Note that the notation $\| . \|$ stands for second norm of vector or matrix in our work. We first investigate the norm of the matrix $(L_n+B_n)^{-1}$. The matrix $(L_n+B_n)$ is time-varying and its elements reach maximum values when the multi-agent graph is strongly connected. In such case, all the off-diagonal terms in $L_n$ become one, and the diagonal matrix $B_n$ becomes an Identity matrix $I_n$. Hence, according to the matrix norm theory (\cite{Book_HornJohn}), we get
\begin{align}\label{eqn:normMat}
  \|[(L_n+B_n)\otimes I_2]^{-1}\|  \leq \min \{\text{eig}(L_n+I_n)\}=0+1=1 .
\end{align}
Thus, the inequality shown in Equation (\ref{eqn:normRs}) takes shape as
\begin{align}\label{eqn:normRs1}
  \| \mathbf{r}_s \| \leq  1 \| k_1 \mathbf{\xi}_1 + k_2 \mathbf{\xi}_2 \|~.
\end{align}
In relation to the definitions of $\xi_1$ and $\xi_2$, the following intends to establish that the effects of the residue terms $\mathbf{\xi}_1$ and $\mathbf{\xi}_2$ are minor in comparison with that of the error terms $\mathbf{e}_p$ and $\dot{\mathbf{e}}_p$, respectively, and so the gross residual term $\| k_1 \mathbf{\xi}_1 + k_2 \mathbf{\xi}_2 \|$ induces negligible influence in the error dynamics (\ref{eqn:errDynmFin}). In this context, we consider a non-negative residual function as shown below:
\begin{align}\label{eqn:errfuncRes}
f(\mathbf{p}_i,\mathbf{p}_j, \alpha_1)= \frac{\|(\mathbf{p}_i - \mathbf{p}_j)^{\alpha_1} - (\mathbf{p}_i-\mathbf{p}_t)^{\alpha_1} + (\mathbf{p}_j-\mathbf{p}_t)^{\alpha_1} \|}{\|(\mathbf{p}_i-\mathbf{p}_t)^{\alpha_1}+(\mathbf{p}_j-\mathbf{p}_t)^{\alpha_1} \|} ~,
\end{align}
In order to restrict the residual function value $f(\mathbf{p}_i,\mathbf{p}_j, \alpha_1)$ in Equation (\ref{eqn:errfuncRes}) by a small positive quantity $\epsilon_1$, s.t. $f(\mathbf{p}_i,\mathbf{p}_j, \alpha_1) \leq \epsilon_1$, the following inequality involving agents position errors needs to be satisfied:
\begin{align}\label{eqn:ineqPij}\nonumber
& \|(\mathbf{p}_i - \mathbf{p}_j)^{\alpha_1} - (\mathbf{p}_i-\mathbf{p}_t)^{\alpha_1} + (\mathbf{p}_j-\mathbf{p}_t)^{\alpha_1} \| \\
& \leq \epsilon_1 \|(\mathbf{p}_i-\mathbf{p}_t)^{\alpha_1}+(\mathbf{p}_j-\mathbf{p}_t)^{\alpha_1} \|~.
\end{align}
Without loss of generality, we now assume all agent positions are one dimensional as: $p_1,p_2,...,p_n,p_t$ for the sake of analysis. We now define a nonlinear residual function of lower dimension as follows.
\begin{align}\label{eqn:funcResidue}
& \bar{f}(r)=\left\{
\begin{array}{lll}
& \frac{(1-r)^{\alpha}-(1-r^{\alpha})}{1+r^{\alpha}}~~\text{for} ~~r < 1, \\
& 0 ~~\text{for}~~$r=1$, \\
& \frac{(r-1)^{\alpha}-(r^{\alpha}-1)}{1+r^{\alpha}}~~\text{for} ~~r > 1,
\end{array}
\right.
\end{align}
where $ r=\frac{|p_j - p_t|}{|p_i - p_t|} \in (0,B_m)$, $\alpha < 1$; $B_m$ is an upper bound on the ratio $r$ based on the spatial constraints on agent locations due to limited communication and dynamic saturation.

The next lemma \cite{CJ_Li} is stated towards finding an upper bound of the nonlinear function $\bar{f}(r)$.

\noindent {\bf Lemma 1.}  If there exists an odd integer or a ratio of odd integers,  $m>1$, then the following inequalities hold for $a,~b \in \Re$.
\begin{align}\label{ineq:err1}
|a+b|^m \leq 2^{m-1} |a^m+b^m |~,  \\ \label{ineq:err2}
\text{and}~~~ |a-b|^m \leq 2^{m-1} |a^m-b^m |~.
\end{align}
In case of $r<1$, the substitution of $m=\frac{1}{\alpha}$, $a=(1-r)^{\alpha}$ and $b = r^{\alpha}$ in Inequality (\ref{ineq:err1}), we get
\begin{align}\label{ineq:err3}
& (1-r)^{\alpha} + r^{\alpha} \leq 2^{1-\alpha}(1-r+r)^{\alpha} \\
& \text{or},~~ \bar{f}(r) \leq 2^{1-\alpha} -1~.
\end{align}
In case of $r>1$, from the definition of $\bar{f}(r)$, we get
\begin{align}\nonumber
& \frac{(r-1)^{\alpha}-(r^{\alpha}-1)}{1+r^{\alpha}} \leq \frac{(r-1)^{\alpha}-(r^{\alpha}-1)}{r^{\alpha}} \\ \label{ineq:err4}
& \text{or},~~\bar{f}(r) \leq (1 - \frac{1}{r})^{\alpha}-1+\frac{1}{r^{\alpha}}~.
\end{align}
Now, by using Inequalities (\ref{ineq:err1}) and (\ref{ineq:err4}), we achieve
\begin{align}
& (1 - \frac{1}{r})^{\alpha}+\frac{1}{r^{\alpha}} \leq 2^{1-\alpha}(1- \frac{1}{r} + \frac{1}{r}) \\
& \text{or},~~\bar{f}(r) \leq 2^{1-\alpha} -1~.
\end{align}
By choosing the parameter $\alpha = 1- \frac{\ln (\epsilon + 1)}{\ln 2} = 1 - \log_2^{(\epsilon+1)} $, we can assure the nonlinear residual function (\ref{eqn:funcResidue}) to be bounded by a small positive quantity $\epsilon$, i.e. $\bar{f}(r) \leq \epsilon$. 

Thus, the inequality shown in Equation (\ref{eqn:ineqPij}) can be satisfied with a choice of parameter $\alpha_1 = 1 - \log_2^{(\epsilon_1+1)}$, s.t. $f(\mathbf{p}_i,\mathbf{p}_j, \alpha_1) \leq \epsilon_1$. In similar fashion, with another choice of parameter $\alpha_2 = 1 -\log_2^{(\epsilon_2+1)}$, the residual function $f(\dot{\mathbf{p}}_i,\dot{\mathbf{p}}_j, \alpha_2)$ can be bounded by a small positive quantity $\epsilon_2$, s.t. $f(\dot{\mathbf{p}}_i,\dot{\mathbf{p}}_j, \alpha_2) \leq \epsilon_2$. At this point, the residual magnitude according to Equation (\ref{eqn:normRs1}) is claimed to be bounded by 
\begin{align}\label{eqn:ineqRs}
\|  \mathbf{r}_s \| \leq  \| \bar{\epsilon}_1 \mathbf{e}_p^{\alpha_1}+\bar{\epsilon}_2 \dot{\mathbf{e}}_p^{\alpha_2} \| ~,
\end{align}
where $\bar{\epsilon}_1$, $\bar{\epsilon}_2 > 0$ depend on $\epsilon_1$ and $\epsilon_2$. The above analysis shows that $| \frac{\partial V_1}{\partial \mathbf{z}_2} \mathbf{r}_s | \ll |\frac{\partial V_1}{ \partial \mathbf{Z}} \mathbf{F}|$ is possible to maintain with proper choice of controller parameters. Now, considering $| \frac{\partial V_1}{\partial \mathbf{z}_2} \mathbf{r}_s | = \backepsilon |\frac{\partial V_1}{ \partial \mathbf{Z}} \mathbf{F}|$ for $\backepsilon > 0$, Equation (\ref{eqn:dotV1}) can be expressed as 
\begin{align}\label{eqn:DervCand2}
\dot{V}_1 = (1 \pm \backepsilon) \frac{\partial V_1}{ \partial \mathbf{Z}} \mathbf{F}~.  
\end{align}
In Equation (\ref{eqn:DervCand2}), the right side of equation is dominated by the homogeneous part $\mathbf{F}$. To this end, the derivative of Lyapunov candidate function shown in Equation (\ref{eqn:DervCand2}) lead to the fact: $\dot{V_1} \leq  -c_2 V_1^{\gamma}$ where $c_2 > 0$ depends on $c_1$ and $\backepsilon$ and $0 < \gamma < 1$. Hence, it is evident that the effects of the residual terms $\mathbf{\xi}_1,\mathbf{\xi}_2$ in multi-agent error dynamics are negligible, and so the error dynamics (\ref{eqn:errDynmFin}) is asymptotically stable and the tracking errors go to zero during the convergence of the closed-loop system.
\noindent  In a multi-agent network, for every connected pair $(i,~j)$, if the inter-agent relative distance vectors, $\mathbf{p}_i(t) - \mathbf{p}_j(t)$, converge to the desired ones, $\mathbf{P}_i - \mathbf{P}_j$, and the network stays connected throughout the dynamics, then the error vector defined above $\mathbf{e}_p=[(\hat{\mathbf{p}}_1-\mathbf{p}_{l})^T, (\hat{\mathbf{p}}_2-\mathbf{p}_{l})^T,.......]^T \rightarrow \mathbf{0}~\text{as}~t \rightarrow \infty$, which implies the convergence of the closed loop system under the new formation control law (\ref{eqn:controllerCJ}). $~~~~~~~~~~\blacksquare$

\section{Concerns to the Connectivity}\label{sec:connectivity}
\noindent  Incorporating fractional powers in the inter-agent system dynamics adds nonlinearities to the governing equations of motion (\ref{eqn:ModelStSp}). It is to be noted that the fractional powers ($\alpha_1,~\alpha_2$) in the proposed controller (\ref{eqn:ConIPij_new}) involve a common parameter $\tau$. In general, the parameter $\tau$ can be chosen from an open range of $\tau > -\frac{1}{2}$, which also ensures the finite time stability of homogeneous systems \cite{Bhat_FT1}. In our controller (\ref{eqn:ConIPij_new}), we choose $\tau$ from a limited range of $-\frac{1}{2} < \tau \leq 0$ to take care of the dynamic network connectivity. A negative value of $\tau$ generates proper fractional powers ($\alpha_1,~\alpha_2$) in the controller, which facilitates in generating smoother UAV trajectories due to moderate fluctuations in the step sizes of UAV dynamics, by utilizing limited control inputs. The following illustrates the justification of such choice of parameter $\tau$ using a simple example.
\subsection{Special Solution to Nonlinear 2nd order ODE:}
\noindent  Finding an explicit solution to a nonlinear dynamic equation is non-trivial. In the following, we determine the explicit state solutions of a nonlinear fractional order system dynamics similar to (\ref{eqn:ModelStSp}) for special initial conditions.

\noindent  The nonlinear inter-agent dynamics in Equation (\ref{eqn:ModelStSp}), can be realized by analyzing a lower dimensional equivalent system dynamics. Consider the following system:
\begin{align}\label{eqn:SysDyn1D}
\left\{
\begin{array}{l l}
    \dot{x}_1 = x_2 \\
    \dot{x}_2 = - k_1  x_1^{1+2\tau} - k_2  x_2^{\frac{1+2\tau}{1+\tau}}~~~\forall \tau > -\frac{1}{2}~.
\end{array}\right.
\end{align}
The double integrator system dynamics in Equation (\ref{eqn:SysDyn1D}) gives rise to the nonlinear 2nd order ODE as
\begin{align}\label{eqn:nonlin2ndODE}
\ddot{x}_1 + k_2  \dot{x}_1^{\frac{1+2\tau}{1+\tau}} + k_1  x_1^{1+2\tau} =0~.
\end{align}
\noindent  We first assume the solution of Equation (\ref{eqn:nonlin2ndODE}) in general form \cite{Hakima_Thesis}, as follows:
\begin{align}\label{eqn:solnGenX1}
x_1(t)=  (x_1^{- \tau}(0) + \tau t)^{-\frac{1}{\tau}}~,
\end{align}
where $x_1(0)$ is the initial condition for state $x_1(t)$. Taking derivatives on both sides of the above equation yields
\begin{align}\label{eqn:solnGenX2}
x_2(t)= \dot{x}_1(t) = - (x_1^{- \tau}(0) + \tau t)^{-\frac{1+\tau}{\tau}}.
\end{align}
Using the expression in Equation (\ref{eqn:solnGenX2}), we derive a relationship between the initial states ($x_1,~x_2$ at $t=0$) in relation to the constant $\tau$, as follows.
\begin{align}\label{eqn:specialInCond}
x_2(0)= - x_1^{1+\tau}(0).
\end{align}
The relationship in Equation (\ref{eqn:specialInCond}) on initial conditions is exploited to determine the special case solution of the nonlinear ODE.
Now, using the system dynamics (\ref{eqn:SysDyn1D}) and Equations (\ref{eqn:solnGenX1}) and (\ref{eqn:solnGenX2}), we obtain
\begin{align}\label{eqn:subst1D}
& \dot{x}_2(t)=  (1+\tau) (x_1^{- \tau}(0) + \tau t)^{-\frac{1+2 \tau}{\tau}} \\ \nonumber
& -k_1 (x_1^{- \tau}(0) + \tau t)^{-\frac{1+2\tau}{\tau}} - k_2 (-1)^{-\frac{1+2\tau}{1+\tau}} (x_1^{- \tau}(0) + \tau t)^{-\frac{1+2\tau}{\tau}}~.
\end{align}
Canceling out the term $(x_1^{- \tau}(0) + \tau t)^{-\frac{1+2\tau}{\tau}}$ from both sides of Equation (\ref{eqn:subst1D}) leads to the following relation.
\begin{align}\label{eqn:consRel1D}
1+\tau = - k_1 - k_2 (-1)^{-\frac{1+2\tau}{1+\tau}}~.
\end{align}
The relation in Equation (\ref{eqn:consRel1D}) involving controller parameters ($\tau,~k_1,~k_2$) can be utilized to obtain a set of gains ($k_1,~k_2$) for a particular value of $\tau$. Thus, the nonlinear state solutions with special initial condition in Equation (\ref{eqn:specialInCond}) can be shown as
\begin{align}\label{eqn:soln1D}
\left\{
\begin{array}{l l}
    x_1(t) =  (x_1^{- \tau}(0) + \tau t)^{-\frac{1}{\tau}}~~~~~;~\tau \neq 0 \\
    x_2(t) = -(x_1^{- \tau}(0) + \tau t)^{-\frac{1+\tau}{\tau}}~.
\end{array}\right.
\end{align}
Equation (\ref{eqn:soln1D}) reveals the correlation between the system states as $|x_1(t)|^{1+\tau}=|x_2(t)|$, which indicates that the rate of convergence of the relative distance $x_1$ is higher than that of the relative velocity $x_2$ for negative $\tau$, and the scenario is opposite for positive $\tau$. This characteristics of the nonlinear system dynamics in Equation (\ref{eqn:SysDyn1D}), is exploited in the proposed controller shown in Equation (\ref{eqn:controllerCJ}), in order to take care of the dynamic connectivity of multi-agent system. An application of the nonlinear controller of Equation (\ref{eqn:controllerCJ}) with a parameter range of $-0.5 < \tau < 0$ facilitates in generating comparatively slower movements in the multi-agent dynamics due to gradual changes in agent velocities, which in turn helps in improving the connectivity of the overall system. It is to be noted that the substitution of $\tau=0$ in Equation (\ref{eqn:SysDyn1D}) gives rise to a second order linear dynamics.

\noindent It can be inferred from the solution in Equation (\ref{eqn:soln1D}) that for a negative $\tau$, both states reach zero (exactly touch zero) at the same time, i.e. $t= -\frac{1}{\tau x_1^{ \tau}(0)}$. In the following, we illustrate the nature of the nonlinear dynamics in Equation (\ref{eqn:SysDyn1D}) using an example.

\begin{figure}
  \centering
   \includegraphics[width=8.8cm,height=6.5cm]{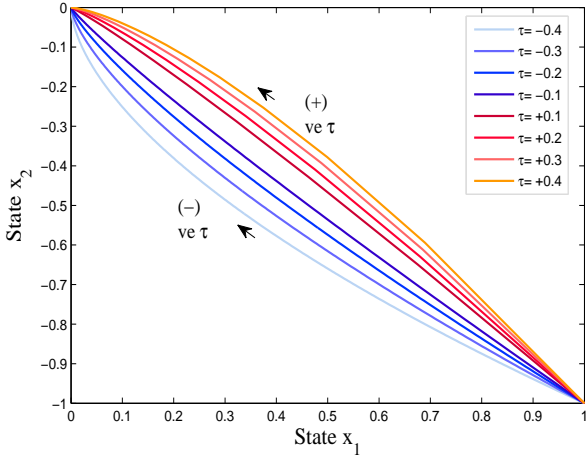}\\
  \caption{Influence of $\tau$ on the state convergence: $\tau \in [-0.4,0.4]-\{0\}$ for the initial condition (1,-1).}\label{fig:numExmp}
\end{figure}

\noindent {\it Numerical example:} Here, we consider $\tau=-\frac{2}{5}$, and the initial condition as $x_1(0)=1$, which implies $x_2(0)=-1$, following the special condition in Equation (\ref{eqn:specialInCond}). The relation involving all the controller parameters become as $1-\frac{2}{5}= -k_1 + k_2$. Satisfying this relation, we select the controller gains as $k_1=-0.3,~k_2=0.3$. The solution states in this case are
\begin{align}\label{eqn:solnExmp1D}
\left\{
\begin{array}{l l}
    x_1(t) =  (1 - \frac{2}{5} t)^{\frac{5}{2}} \\
    x_2(t) = -(1 - \frac{2}{5} t)^{\frac{3}{2}}~.
\end{array}\right.
\end{align}
In this example, the system states are related by $|x_1(t)|^3=|x_2(t)|^5$, indicating the faster convergence rate of state $x_1$. The effect of the parameter $\tau$ on the state convergence speed is depicted in Fig. \ref{fig:numExmp}.
Not only in the controller, fractional powers can also be used to obtain fast estimates, if used in a nonlinear observer as explained in the following.

\section{Nonlinear State Observer}
In this section, we illustrate a nonlinear observer devised to obtain the velocity estimate of a UAV using its position coordinates and heading angles. The position coordinates and heading angles are assumed to be available or known. To develop the observer, we first transform the original states as shown below:
\begin{align}\label{eqn:stateTrnsfmEstm}
& \left\{
\begin{array}{lll}
z_i = x_i \cos(\phi_i) + y_i \cos(\phi_i) \\
w_i = y_i \cos(\phi_i) - x_i \sin(\phi_i) \mathbf{u}_i[2] .
\end{array}
\right.
\end{align}
Suppose that the estimate of $i$th agent velocity, $v_i$, is denoted by $\hat{v}_i$, and the error of estimation is: $e_i = v_i - \hat{v}_i$. The nonlinear observer equations are given as
\begin{align}\label{eqn:nonlnrObs}
& \left\{
\begin{array}{lll}
\dot{\hat{z}}_i = \hat{v}_i + w_i + sgn(e_i)  e_i^{3/5} \\
\dot{\hat{w}}_i = \mathbf{u}_i[1] + sgn(e_i) abs(e_i)^{1/5} .
\end{array}
\right.
\end{align}
The observer in Equation (\ref{eqn:nonlnrObs}) generates an estimate of $v_i$ using the known inputs of $x_i$, $y_i$ and $\phi_i$. The error term in Equation (\ref{eqn:nonlnrObs}) has fractional powers (\cite{obsvCJ}), which are incorporated to enhance the convergence speed of the estimation process.
\subsection{Example:}
Here, we present a simple example to demonstrate the effectiveness of the nonlinear observer over the linear one. Consider a linear system dynamics governed by the following equation.
\begin{align}\label{eqn:exmpDynm}
& \left\{
\begin{array}{lll}
\dot{y}_1 = y_2 \\
\dot{y}_2 = -y_1 -y_2~,
\end{array}
\right.
\end{align}
where, $y_1$ and $y_2$ are the system states, $y_1$ is the output. Suppose the state estimates of $y_1$ and $y_2$ are $\hat{y}_1$ and $\hat{y}_2$, respectively. The traditional linear observer for estimating the output is given as
\begin{align}\label{eqn:linObsv}
& \left\{
\begin{array}{lll}
\hat{\dot{y}}_1 = \hat{y}_2 + (y_1 - \hat{y}_1)  \\
\hat{\dot{y}}_2 = - \hat{y}_1 - \hat{y}_2 + (y_1 - \hat{y}_1) .
\end{array}
\right.
\end{align}
The nonlinear state observer is given as
\begin{align}\label{eqn:nonlinObsv}
& \left\{
\begin{array}{lll}
\hat{\dot{y}}_1 = \hat{y}_2 + sgn(y_1 - \hat{y}_1) . (abs(y_1 - \hat{y}_1))^{3/5}  \\
\hat{\dot{y}}_2 = - \hat{y}_1 - \hat{y}_2 + sgn(y_1 - \hat{y}_1) . (abs(y_1 - \hat{y}_1))^{1/5} .
\end{array}
\right.
\end{align}
The performance of the linear observer in Equation (\ref{eqn:linObsv}) and the nonlinear observer in Equation (\ref{eqn:nonlinObsv}), are shown in Fig. $4$. The assumed initial states and estimates are $[1,-1]$ and $[-1,1]$, respectively. Fig. $4$ exhibits that the nonlinear observer results in faster convergence of the state estimate as compared to that of the linear observer. More details about the nonlinear observer with fractional powers are available in \cite{obsvCJ}.
\begin{figure}[H]
\label{fig:ObsExmp}
  \centering
   \includegraphics[width=8.2cm,height=6cm]{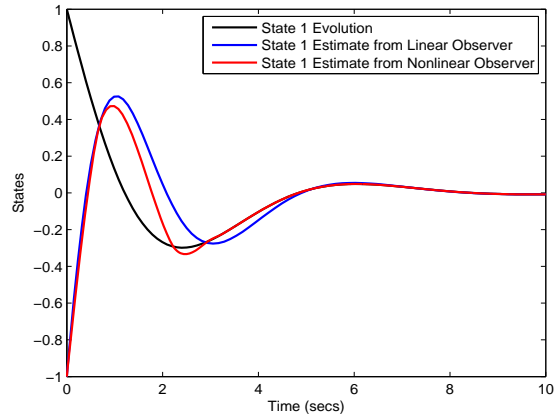}\\
  \caption{Estimation Performance by Linear and Nonlinear Observers.}
\end{figure}

\section{Simulation Results}\label{sec:results}
In simulation, we consider a network of four UAVs trying to make a target-centric formation. The simulation parameters are itemized as follows:
\begin{itemize}
  \item Desired distance from the target: $\delta=100$ m
  \item Desired angle of UAV w.r.t. the target: $\psi_i=\frac{2\pi i}{3}$
  \item UAV velocity bounds: $v_{min}=5$ m/s, and $v_{max}=25$ m/s
  \item Target velocity: $v_{l}=10$ m/s, and initial heading: $\psi_t=0$ rad
  \item Communication parameters: range $R=300$ m, slope $\sigma=10$
\end{itemize}
\begin{table}[h]
\footnotesize
\caption{Initial States of the UAVs}
\label{tab:initcond}\centering
\begin{tabular}{c c c c c c}
  \hline
   Set & UAV & X co (m) & Y co (m) & $v_i$ (m/s) & $\phi_i$ (rad) \\ \hline \\[-2mm]
   & 1 &  18.2249 & 71.4778 & 8 & 0 \\
   A & 2 &  -11.6509 &  97.6854 & 8.5 &  0.7854 \\
   & 3 &  -1.4301 & 133.4849 & 9 & 1.5708 \\
   & 4 & 3.8123 & 103.1000 & 9.5 & 2.3562 \\
  \hline  \\[-2mm]
   & 1 & -12.2025  & -13.1759 & 5 & 0 \\
   B & 2 & -35.1523  & 109.6072  & 5.5 &  0.7854 \\
   & 3 & 131.2880  & 89.3857 & 6 & 1.5708 \\
   & 4 & 65.2199 & 134.8779 & 6.5 & 2.3562 \\
  \hline \hline
\end{tabular}
\end{table}
The target starts from the origin and maneuvers in a sinusoidal path with acceleration input $[0~~0.5 \sin(\frac{2\pi t}{50})]^T$. Two sets of initial conditions of UAVs, used in the simulation, are shown in Table \ref{tab:initcond}. For initial condition A, the initial network connectivity is greater than the final desired one, and for initial condition B, the scenario is exactly opposite.\\\\
\noindent  We now state the performance of the proposed observer in estimating every agent's velocity using its position and heading information. In simulation, we consider the UAV initial states enlisted in Table \ref{tab:initcond}. The controller parameters are selected as $k_1=1,k_2=1.6$  and  $\tau= -0.1$. Fig. $5$ shows the original state and estimate evolutions over time during the formation dynamics.\\\\
\noindent  We use fixed controller gains in the simulations as $k_1=1,~k_2=1.6$, and consider different cases (Cases 1-5) with varying parameter $\tau$ as $\tau= -0.2,-0.1,0, 0.1, 0.2$ in the controller. The results of the proposed controller for initial condition A, are shown in Figs. $6$-$10$. The formation convergence speed increases (\cite{Bhat_FT1}) as the value of $\tau$ increases; at the same time results also show that a faster dynamics with high $\tau$ value may have adverse effects on the network connectivity profile. A low value of $\tau$ reduces the control input requirements of linear acceleration and angular velocity, causing slow changes in the agent velocities. Fig. $11$ and $12$ show how the time-varying connectivity profile changes for a range of values of the parameter $\tau$, i.e. $\{ -0.2,-0.1,0, 0.1, 0.2 \}$, for initial conditions A and B, respectively. \vspace*{-1cm}
\begin{figure}[ht]
\label{fig:SimObsv}
  \centering
   \includegraphics[width=9cm,height=6.8cm]{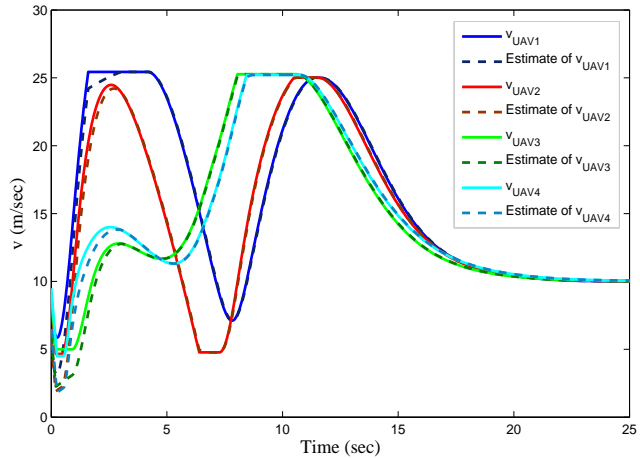}\\
  \caption{Nonlinear Observer Performance in estimating Multi-UAV velocities.}
\end{figure}\vspace*{-3cm}
\subsection{Discussion}\vspace*{-1cm}
The fractional power parameters in the controller provide an additional adjustment capability in response to the time-varying connectivity, the closed-loop convergence speed, the amount of control input used, and the trajectory smoothness. The controller with appropriate parameter adjustments has potential to maintain and improve the time-varying network connectivity during the formation dynamics. For a range of power parameters, the controllers generate a family of trajectories which can be chosen according to the specific requirement during a cooperative mission.\\\\
\noindent  The advantages of using proper fractional powers (for $ - 0.5 < \tau < 0$) over improper powers (for $\tau >0$) are: (1) Generating smooth UAV trajectories causing less ripples in the multi-agent dynamics; (2) Requirement of low control effort to accomplish the formation task; (3) Reducing stiffness of the rises and valleys in the time-varying connectivity profile. The only disadvantage associated with the constrained parameter space is comparatively slower convergence speed of the closed-loop system than that of $\tau > 0$. On the other hand, a positive value of $\tau$ generates improper fractional powers, which results in a higher requirement of the control inputs used by UAVs in order to accomplish the same task.

\begin{figure}[H]
 \centering
 \subfigure[Agents' trajectories in cartesian coordinates]{
  \includegraphics[width=8cm,height=5cm]{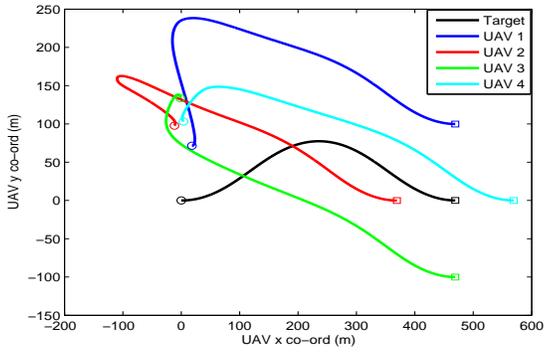}
   \label{fig5}
   }
 \subfigure[Control inputs]{
  \includegraphics[width=8cm,height=5cm]{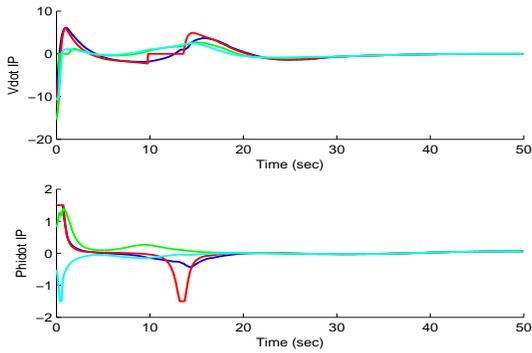}
   \label{fig6}
   }
 \subfigure[Velocity and heading]{
  \includegraphics[width=8cm,height=5cm]{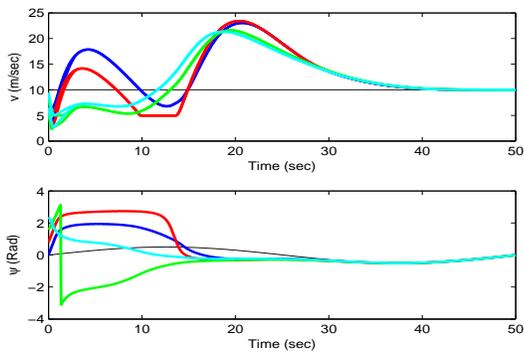}
   \label{fig8}
   }
 \subfigure[Position errors]{
  \includegraphics[width=8cm,height=5cm]{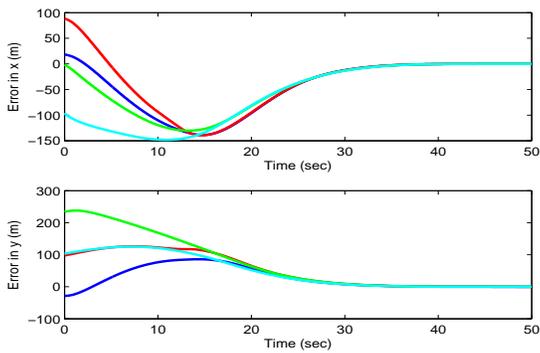}
   \label{fig7}
   }

 \label{fig:set2-}
 \caption{Results of the formation controller with $\tau= -0.2$.}
\end{figure}

\begin{figure}[H]
 \centering
 \subfigure[Agents' trajectories in cartesian coordinates]{
  \includegraphics[width=8cm,height=5cm]{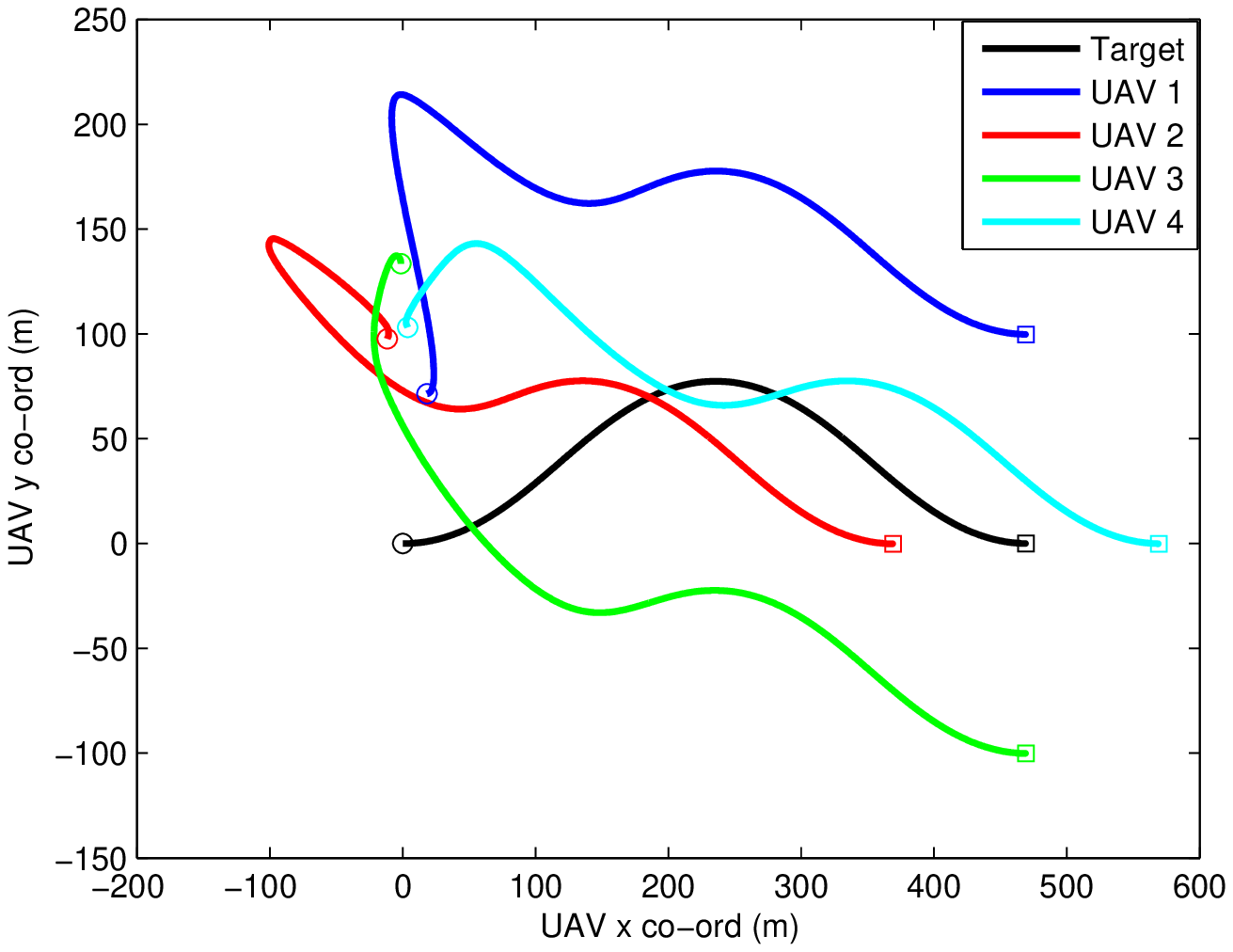}
   \label{fig5}
   }
 \subfigure[Control inputs]{
  \includegraphics[width=8cm,height=5cm]{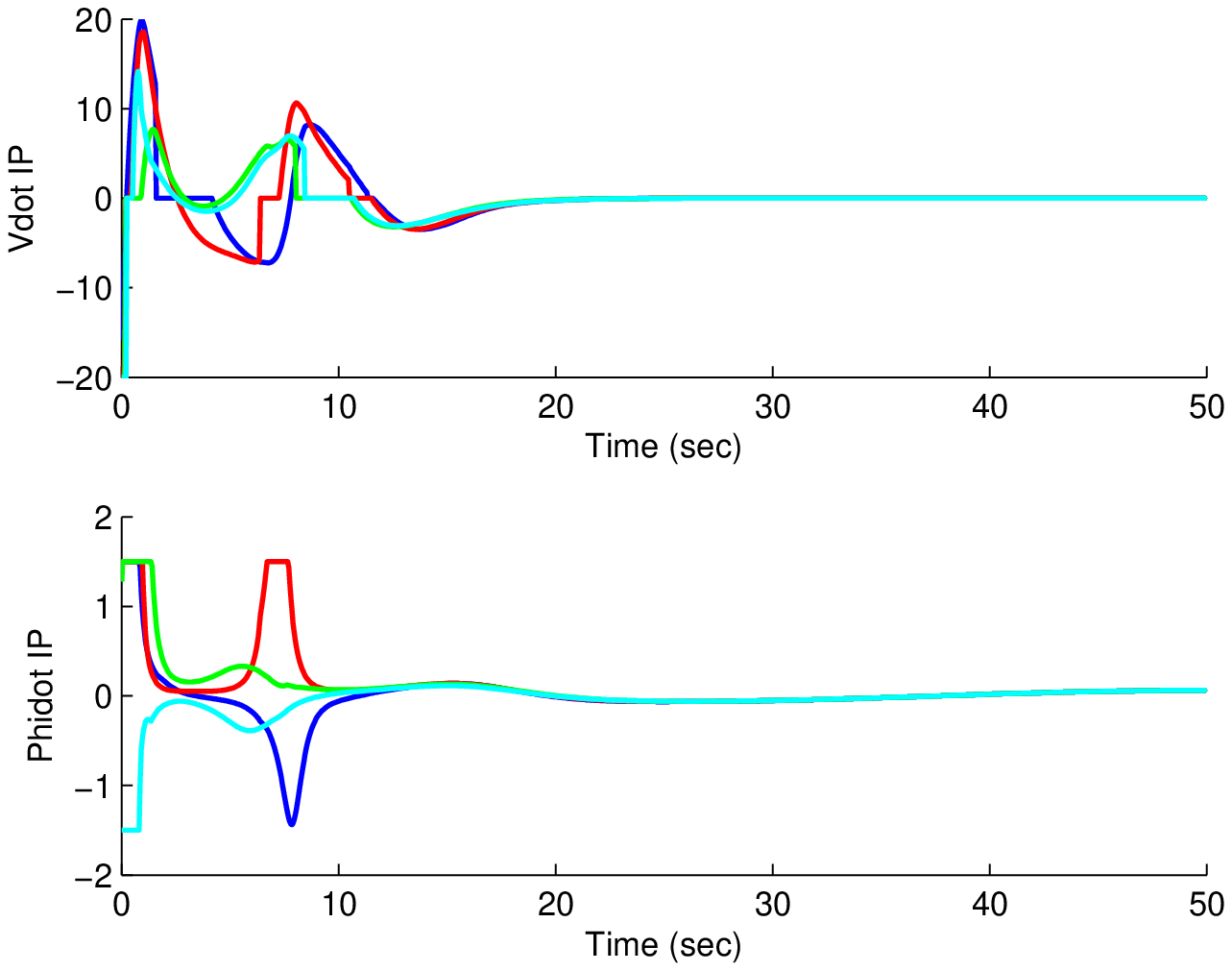}
   \label{fig6}
   }
 \subfigure[Velocity and heading]{
  \includegraphics[width=8cm,height=5cm]{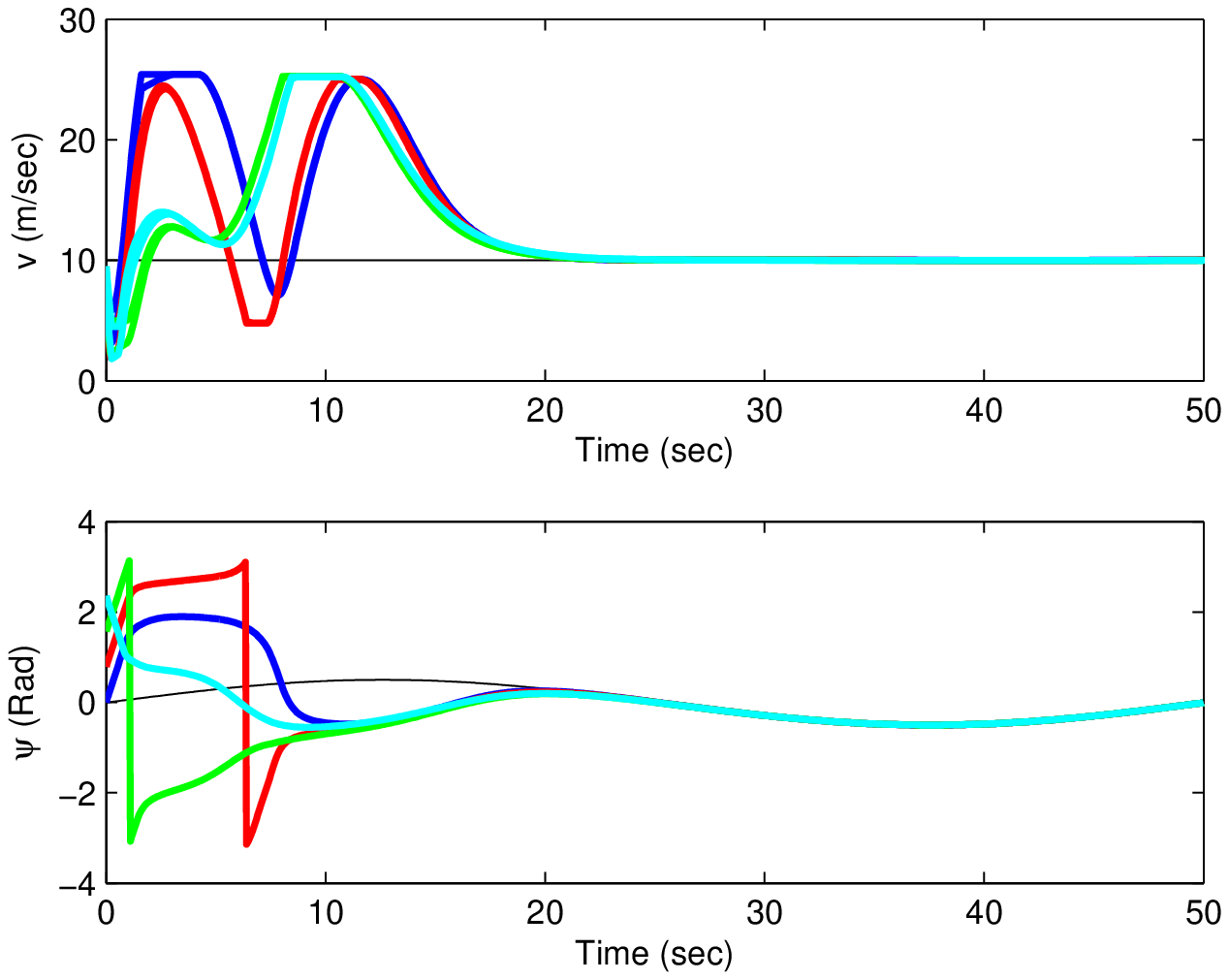}
   \label{fig8}
   }
 \subfigure[Position errors]{
  \includegraphics[width=8cm,height=5cm]{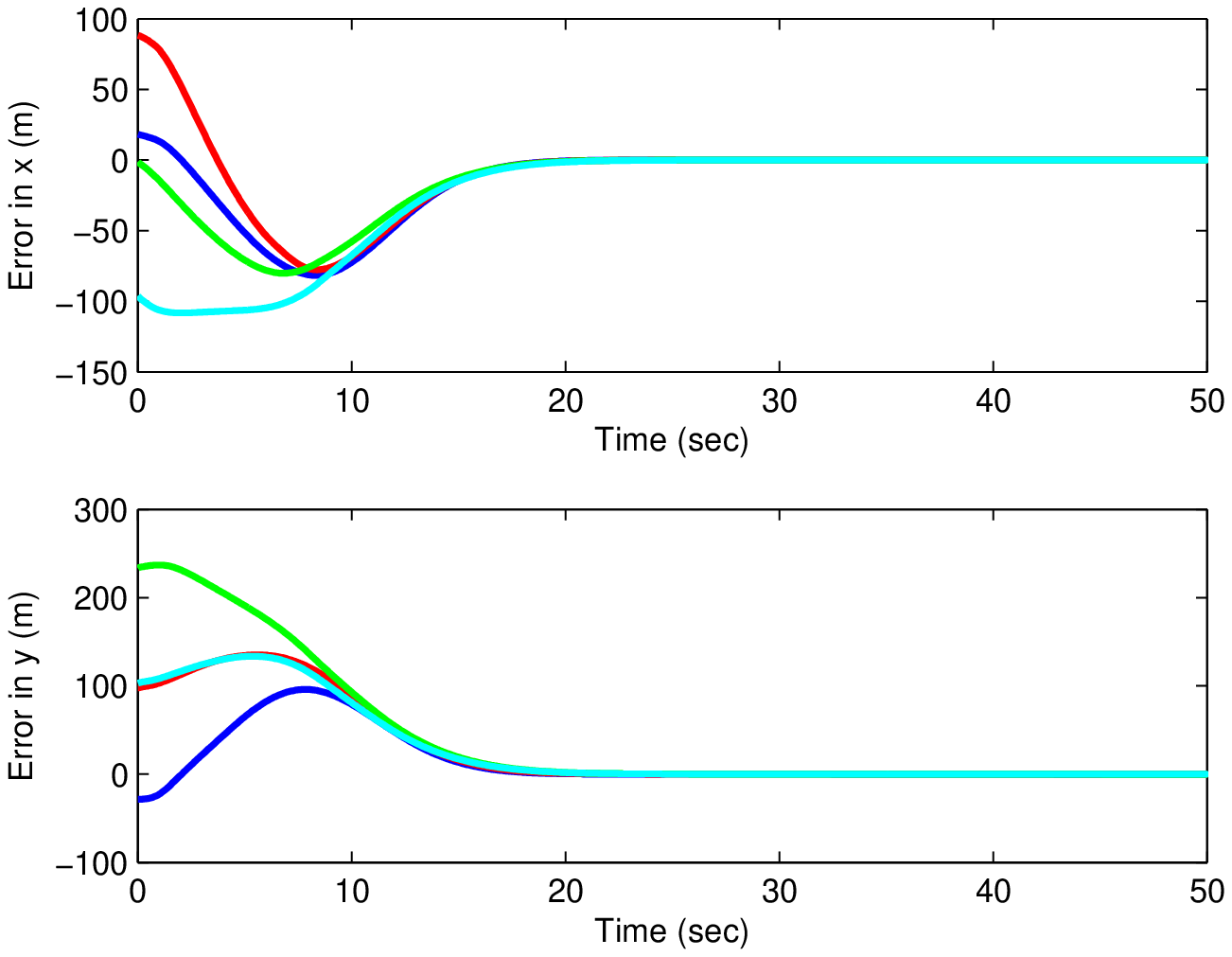}
   \label{fig7}
   }

 \label{fig:set1-}
 \caption{Results of the formation controller with $\tau= -0.1$.}
\end{figure}

\begin{figure}[H]
 \centering
 \subfigure[Agents' trajectories in cartesian coordinates]{
  \includegraphics[width=8cm,height=5cm]{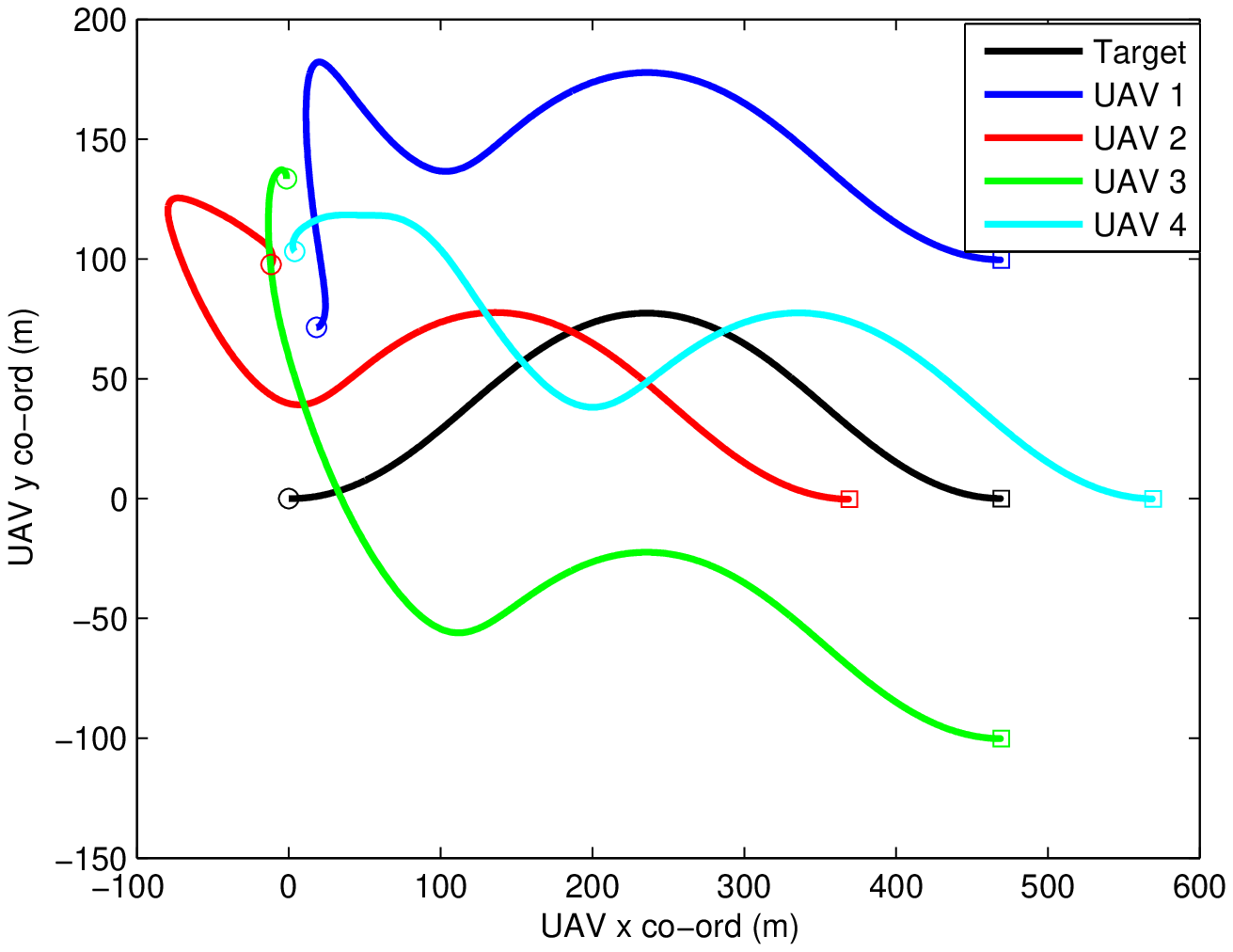}
   \label{fig5}
   }
 \subfigure[Control inputs]{
  \includegraphics[width=8cm,height=5cm]{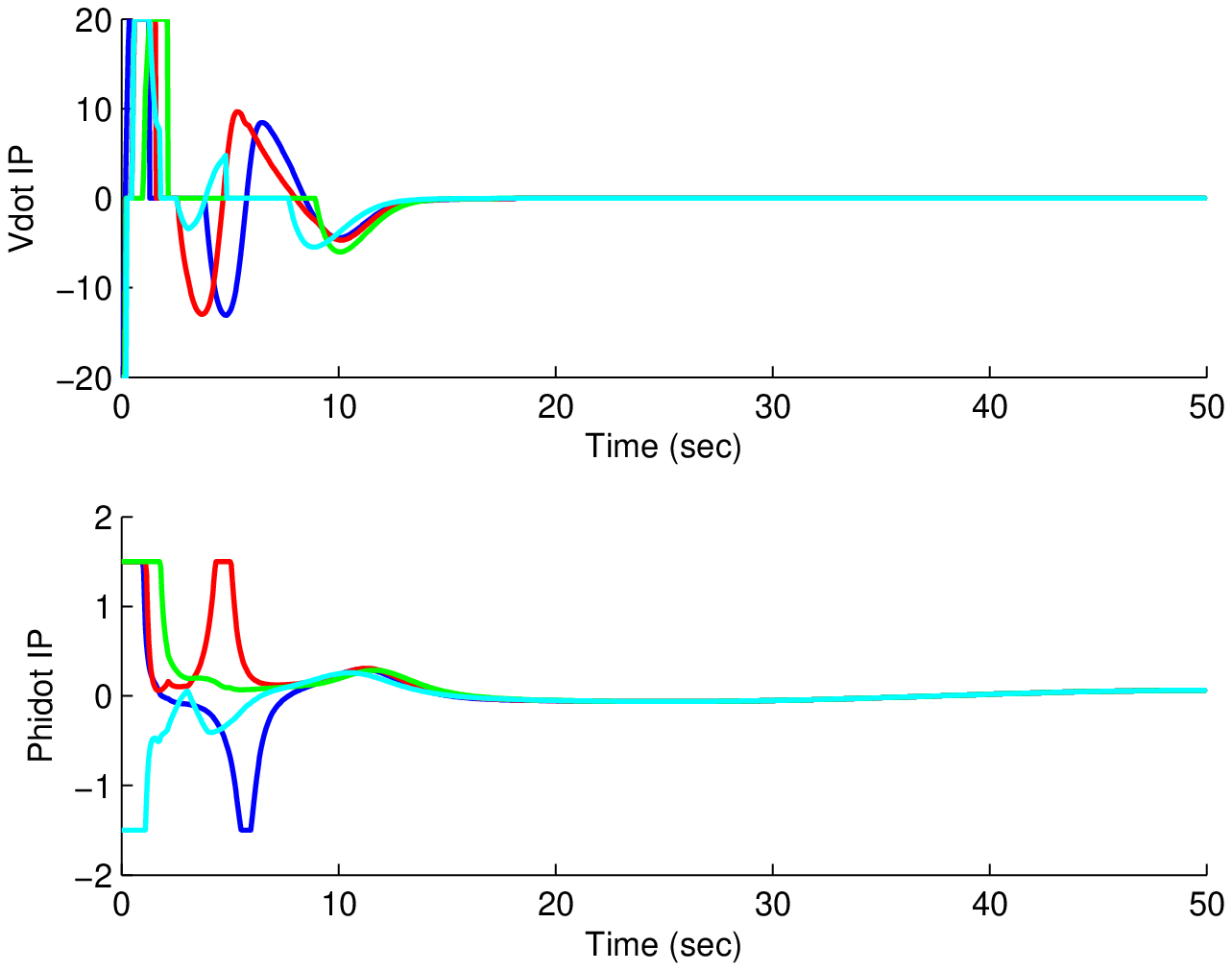}
   \label{fig6}
   }
 \subfigure[Velocity and heading]{
  \includegraphics[width=8cm,height=5cm]{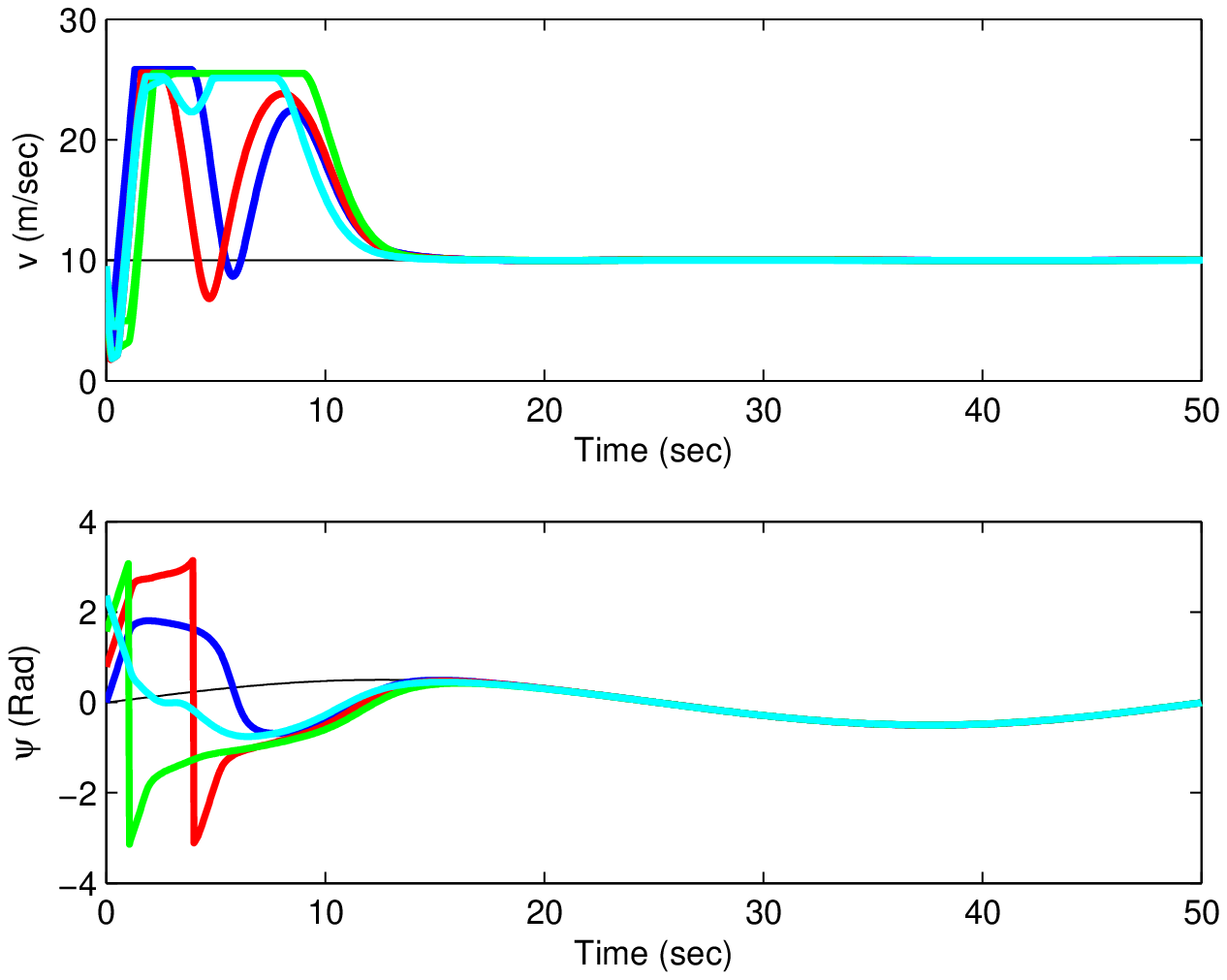}
   \label{fig8}
   }
 \subfigure[Position errors]{
  \includegraphics[width=8cm,height=5cm]{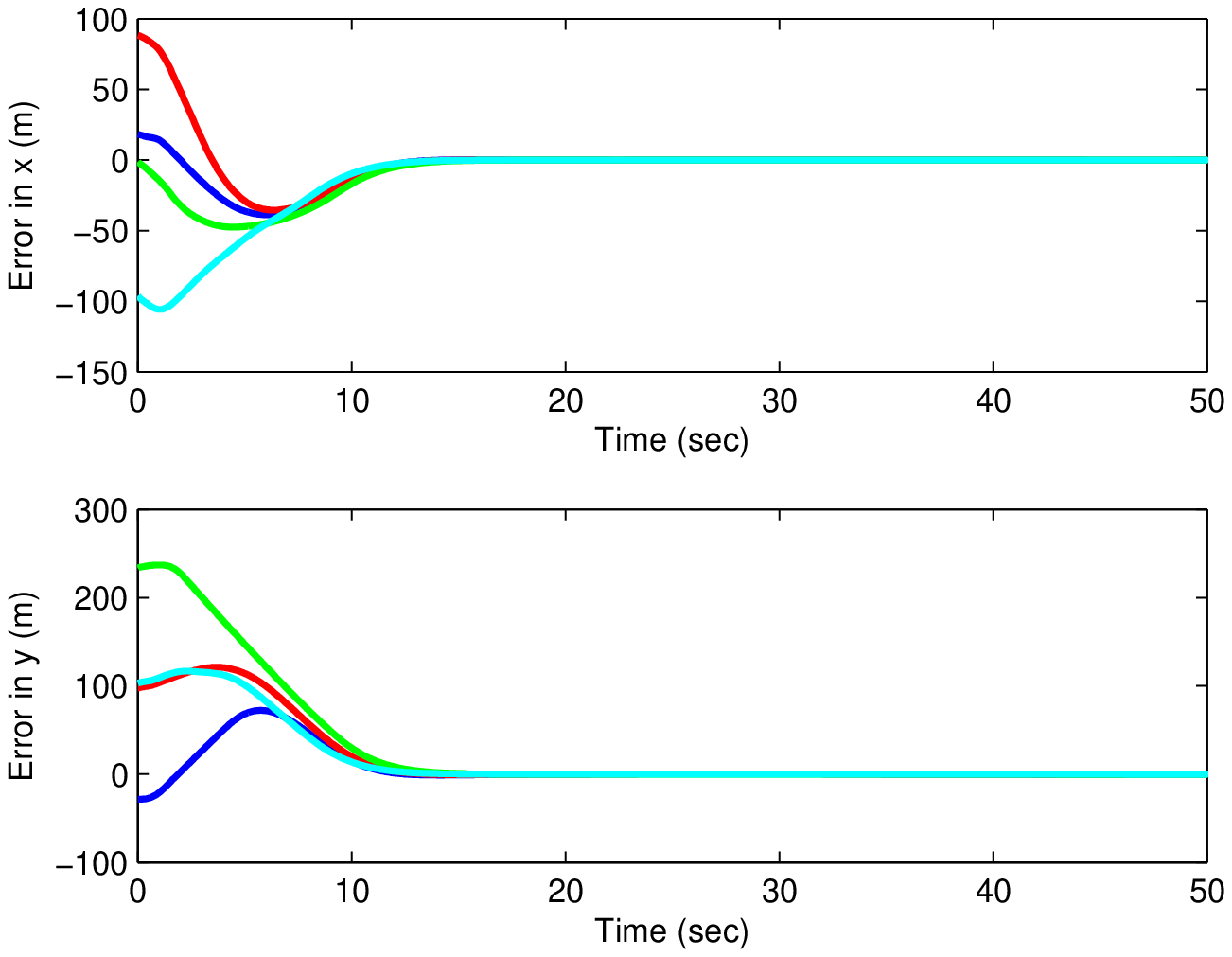}
   \label{fig7}
   }
 \label{fig:set0}
 \caption{Results of the formation controller with $\tau= 0$.}
\end{figure}

\begin{figure}[H]
 \centering
 \subfigure[Agents' trajectories in cartesian coordinates]{
  \includegraphics[width=8cm,height=5cm]{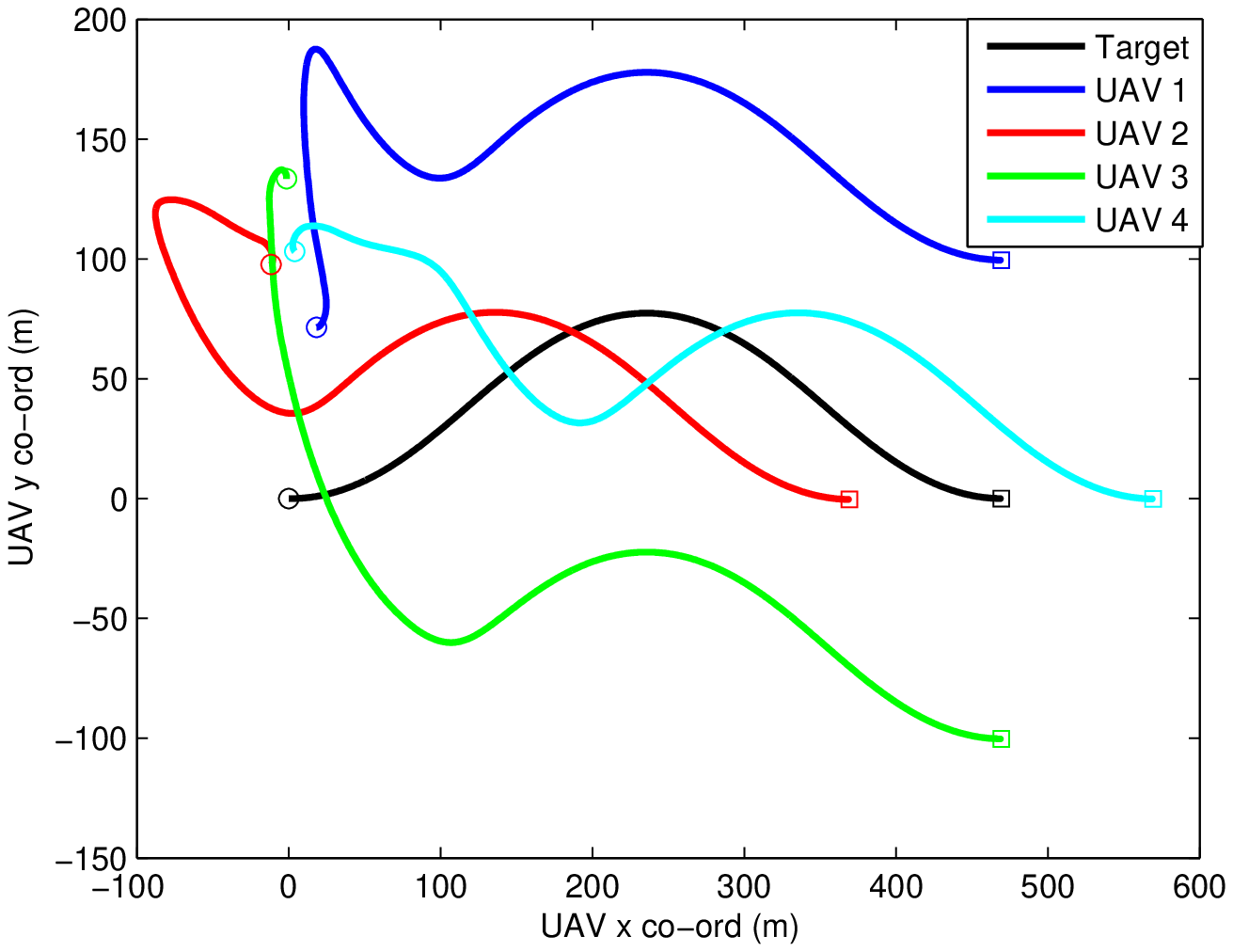}
   \label{fig5}
   }
 \subfigure[Control inputs]{
  \includegraphics[width=8cm,height=5cm]{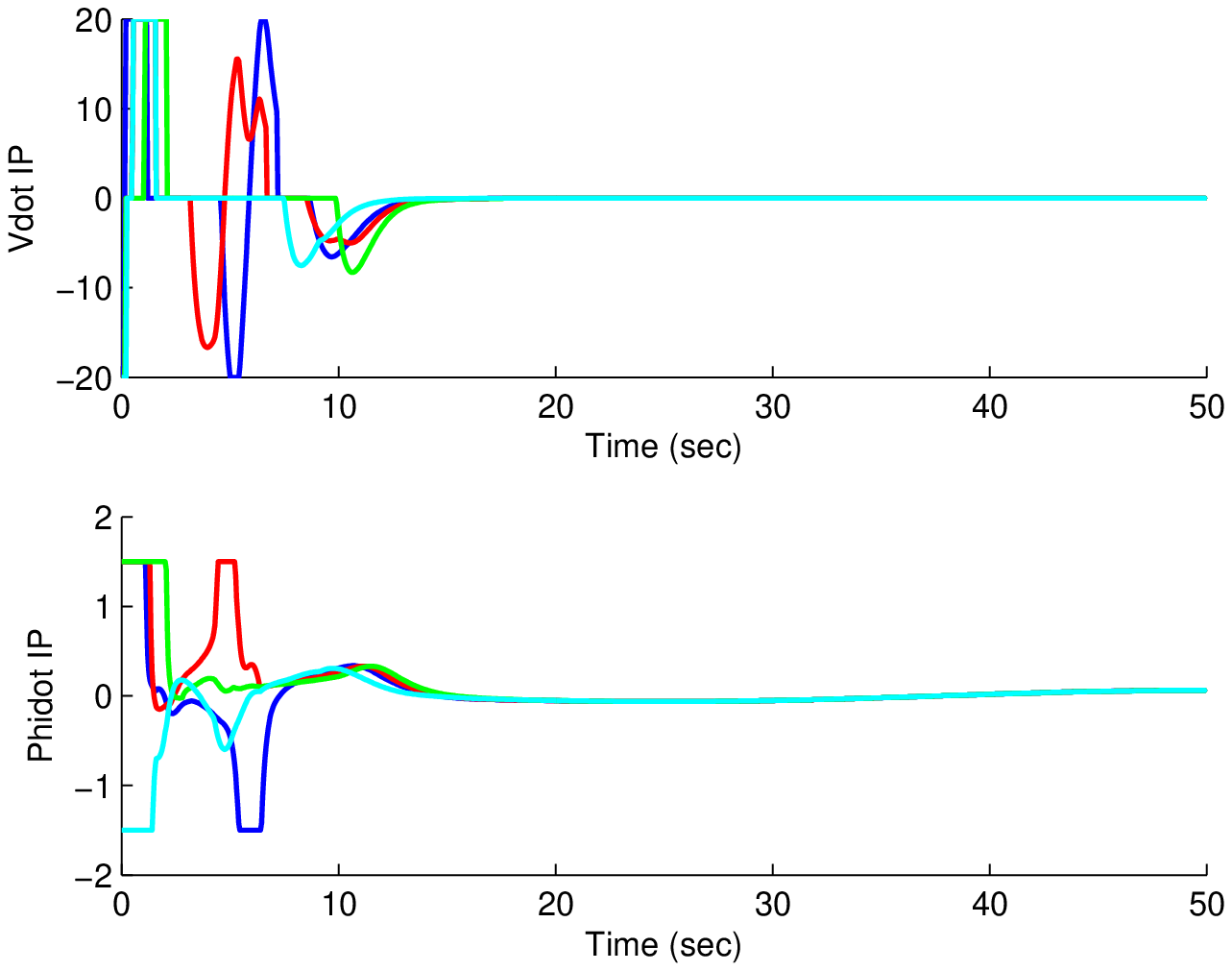}
   \label{fig6}
   }
 \subfigure[Velocity and heading]{
  \includegraphics[width=8cm,height=5cm]{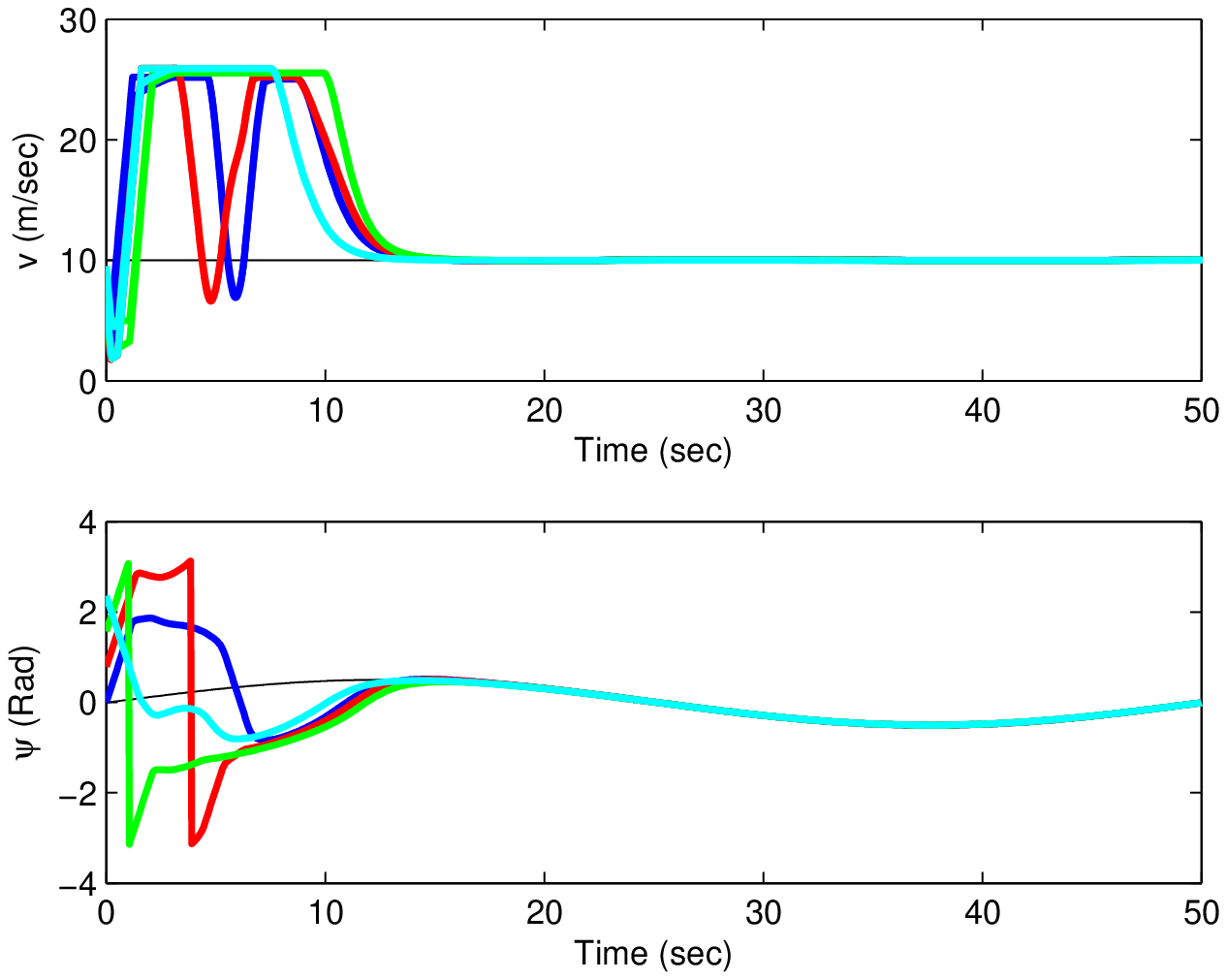}
   \label{fig8}
   }
 \subfigure[Position errors]{
  \includegraphics[width=8cm,height=5cm]{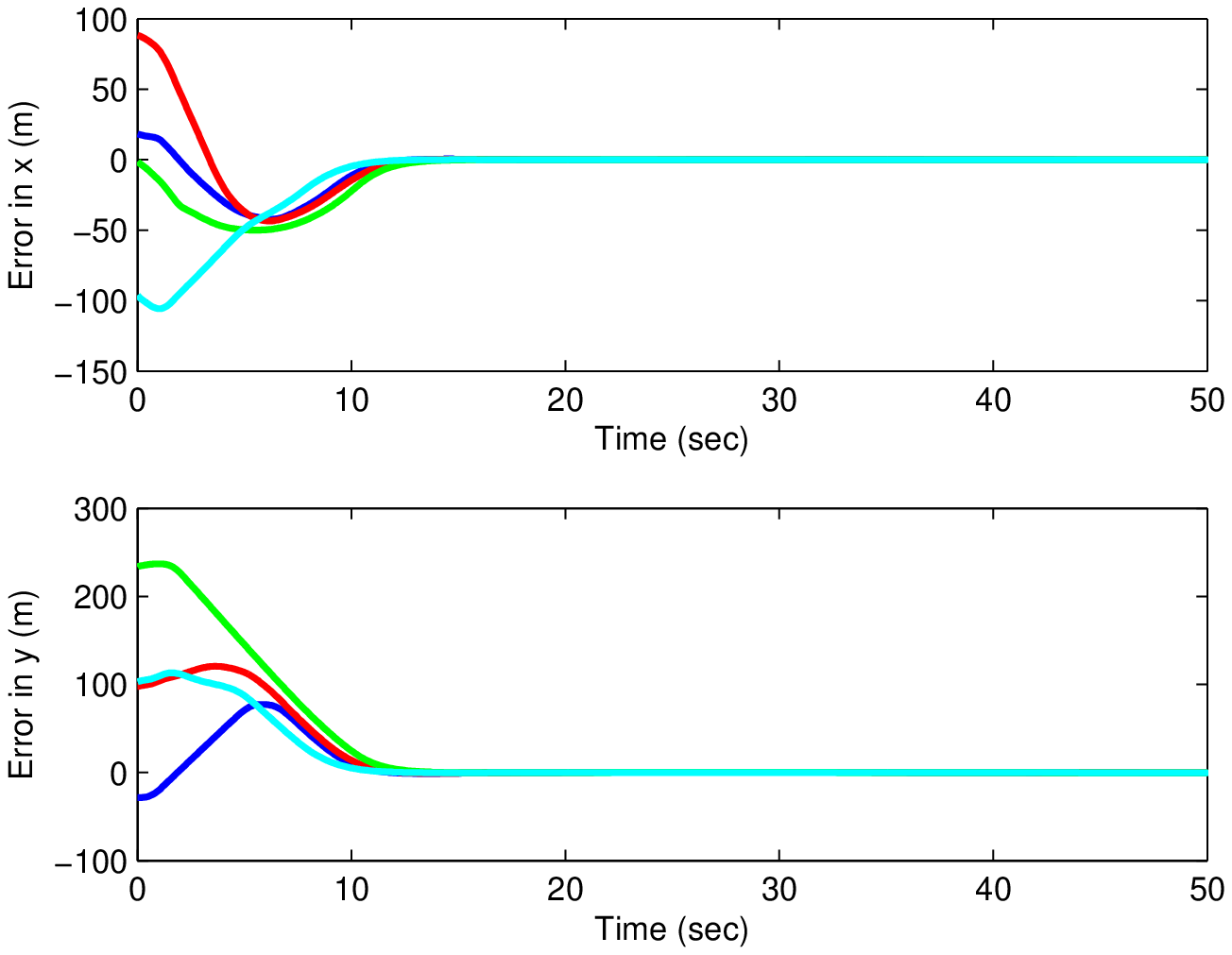}
   \label{fig7}
   }

 \label{fig:set1+}
 \caption{Results of the formation controller with $\tau= 0.1$.}
\end{figure}

\begin{figure}[H]
 \centering
 \subfigure[Agents' trajectories in cartesian coordinates]{
  \includegraphics[width=8cm,height=5cm]{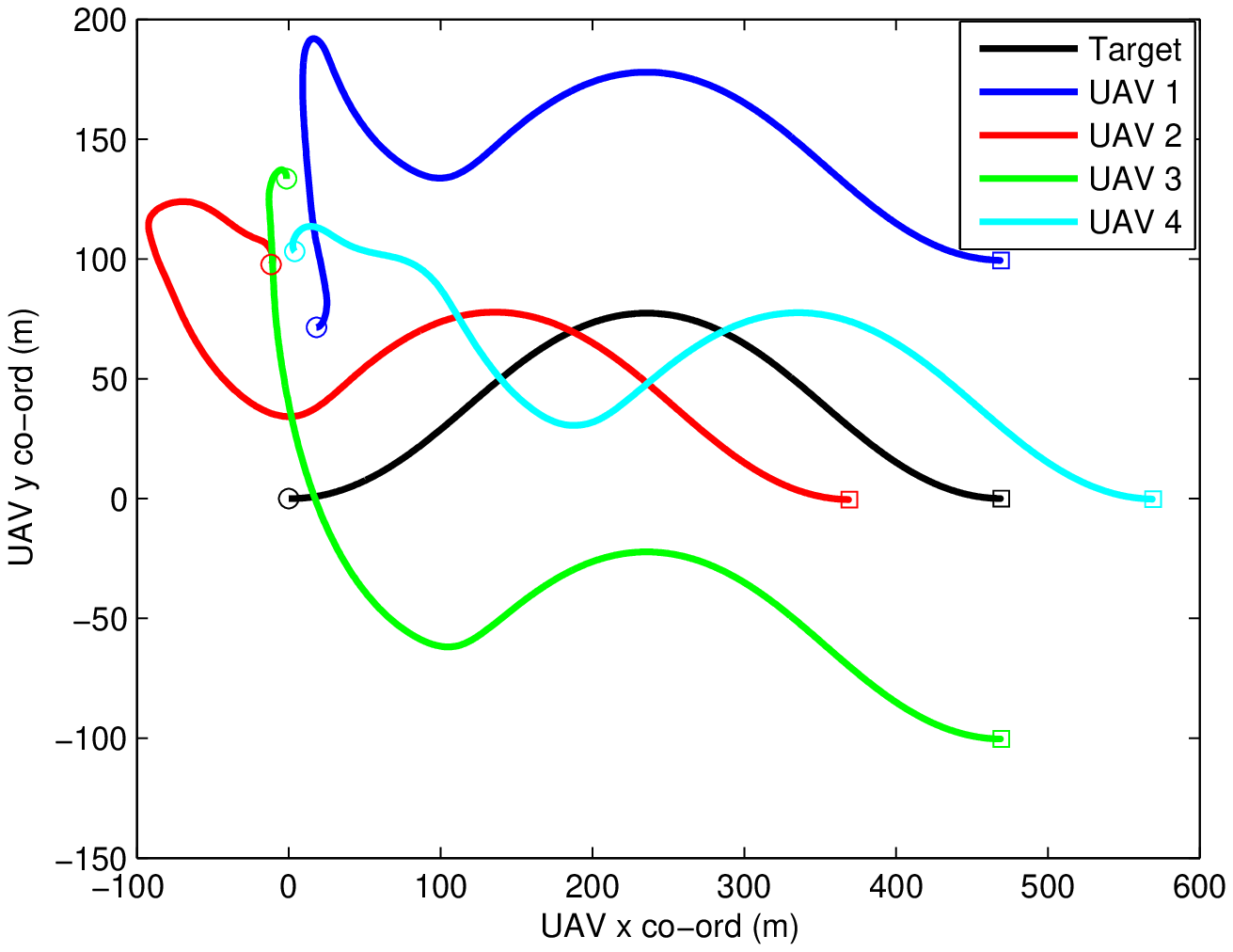}
   \label{fig5}
   }
 \subfigure[Control inputs]{
  \includegraphics[width=8cm,height=5cm]{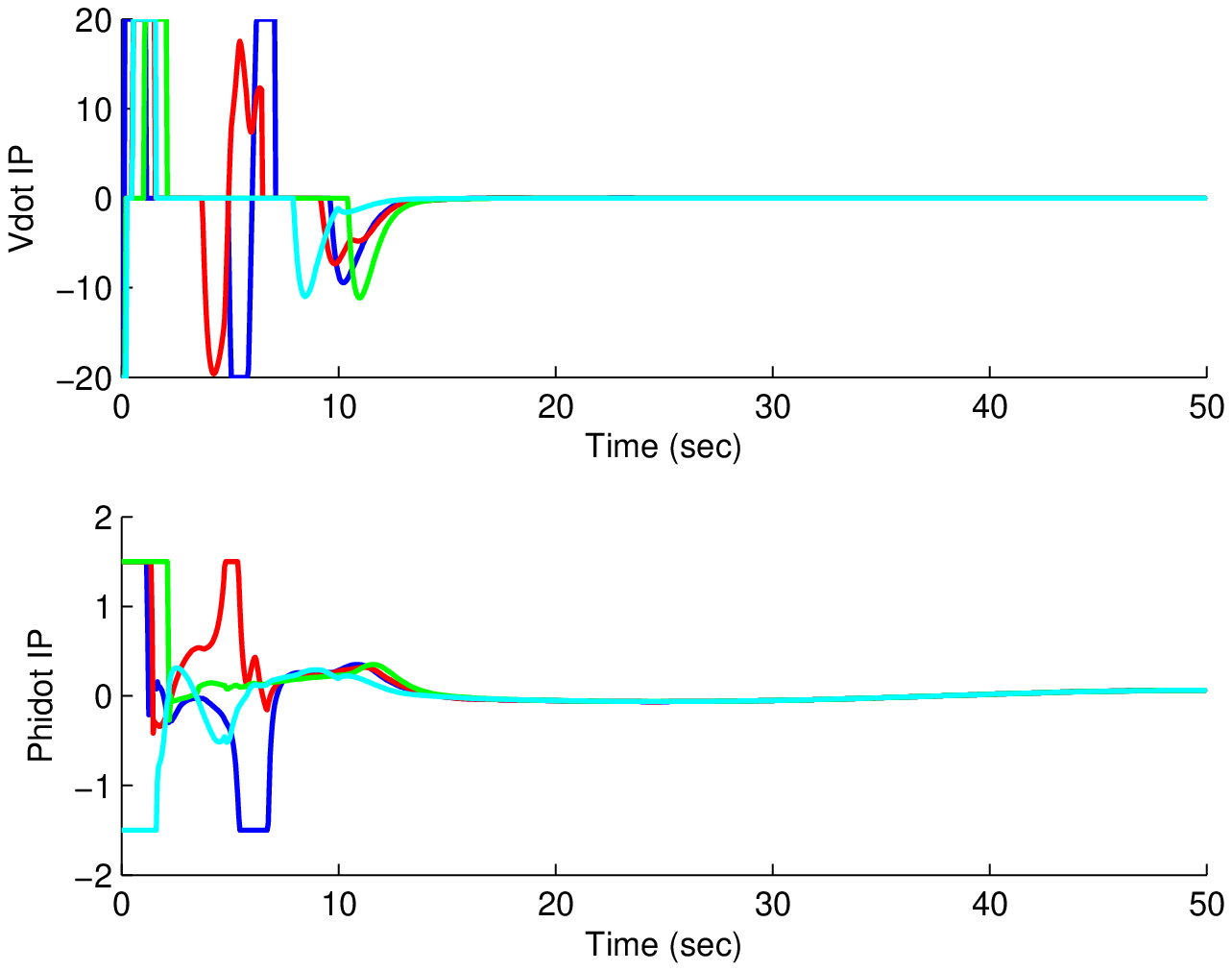}
   \label{fig6}
   }
 \subfigure[Velocity and heading]{
  \includegraphics[width=8cm,height=5cm]{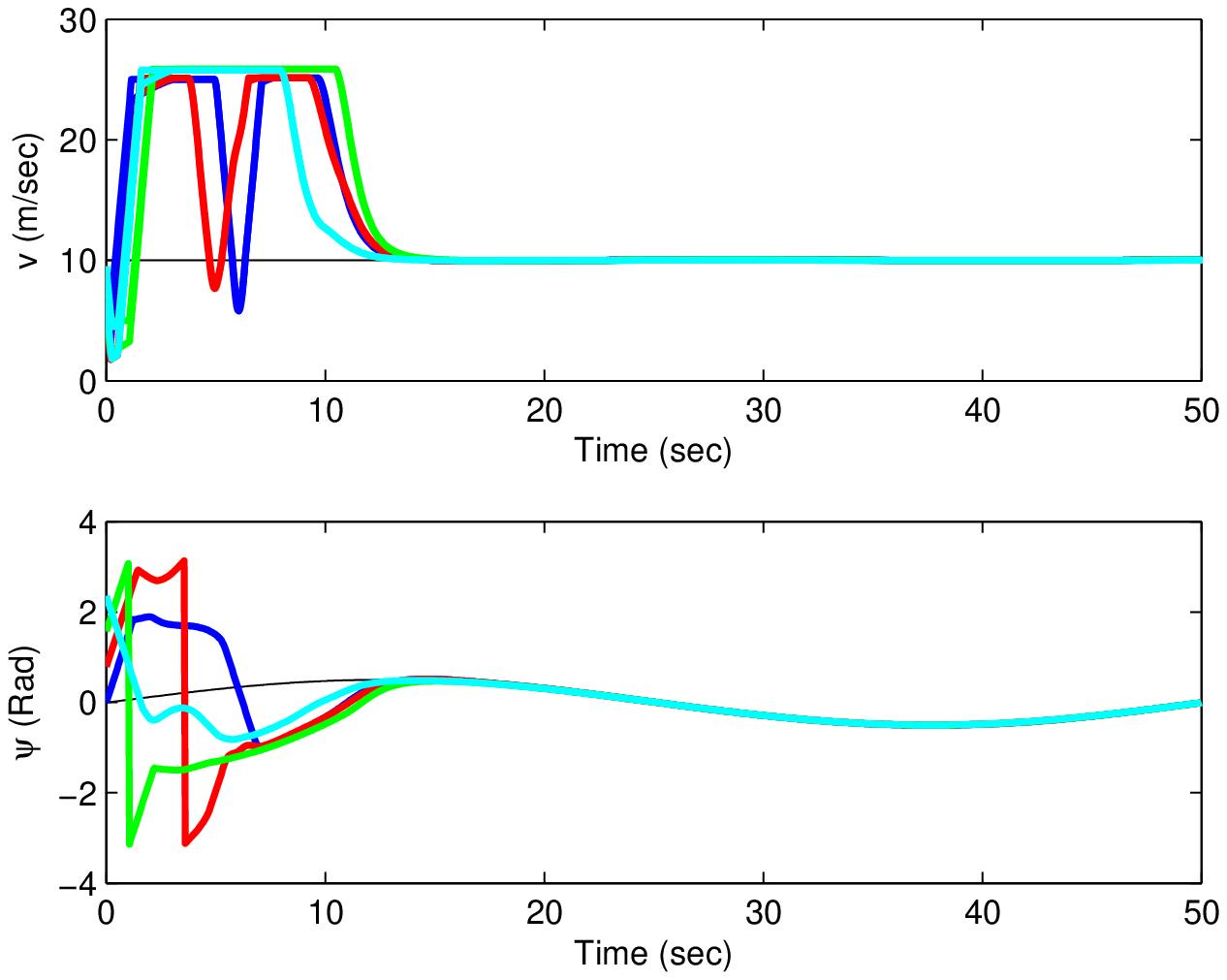}
   \label{fig8}
   }
 \subfigure[Position errors]{
  \includegraphics[width=8cm,height=5cm]{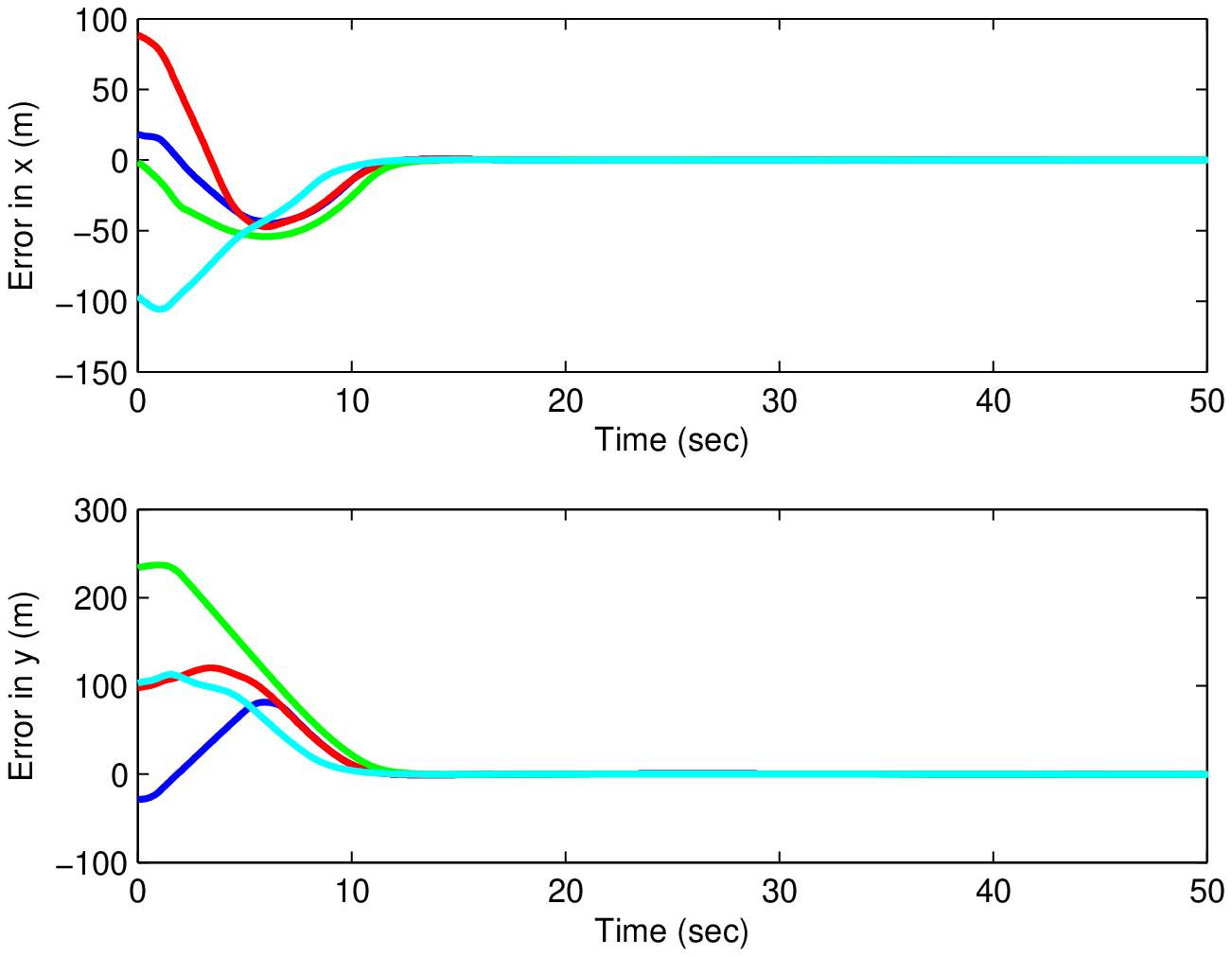}
   \label{fig7}
   }
 \label{fig:set2+}
 \caption{Results of the formation controller with $\tau= 0.2$.}
\end{figure}

\begin{figure}[H]
\label{fig:varyTau}
  \centering
   \includegraphics[width=8.5cm,height=6cm]{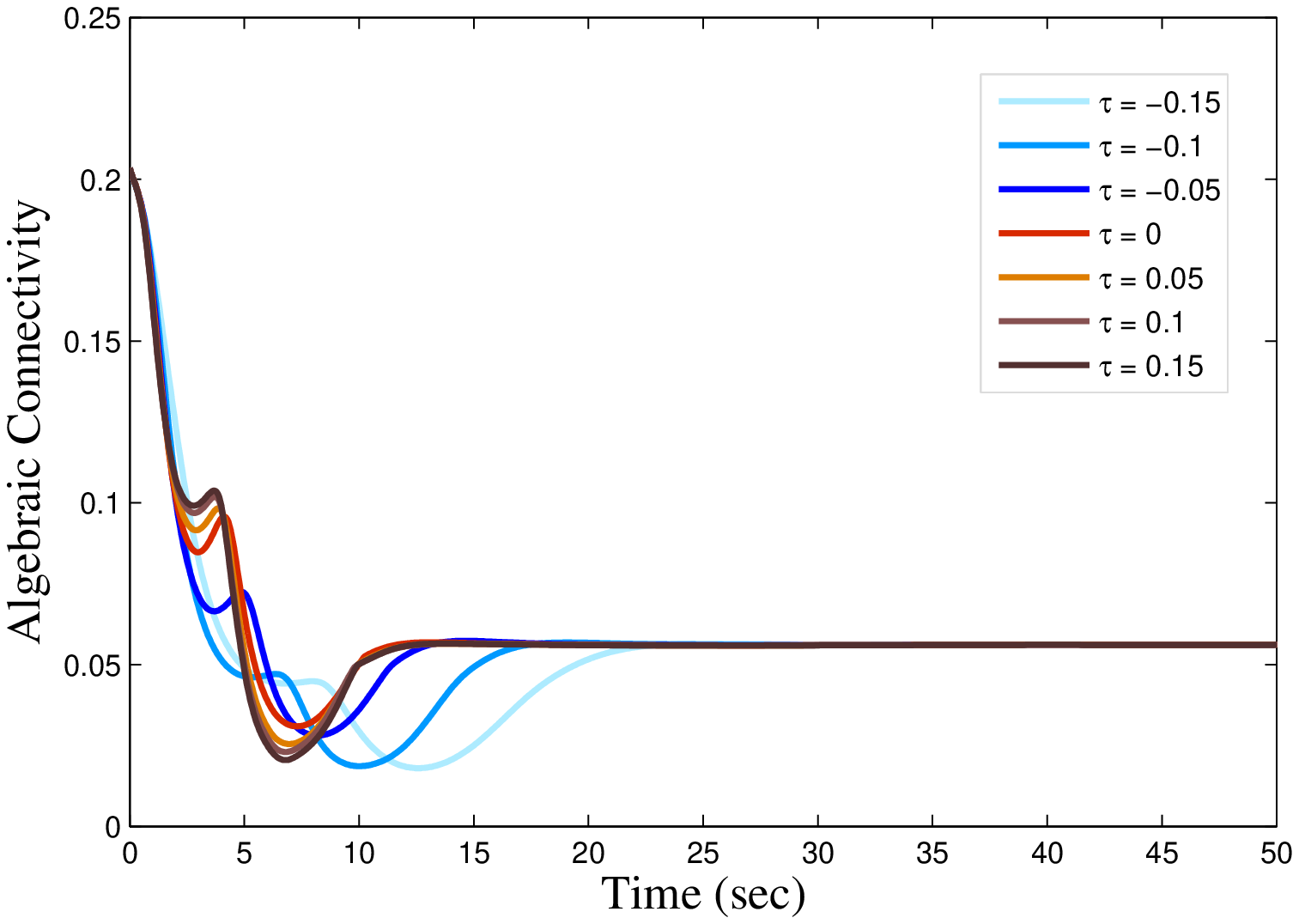}\\
  \caption{Connectivity profiles with varying parameter $\tau$ for initial condition A in Table \ref{tab:initcond}.}
\end{figure}

\begin{figure}[H]
\label{fig:varyTau}
  \centering
   \includegraphics[width=8.5cm,height=6cm]{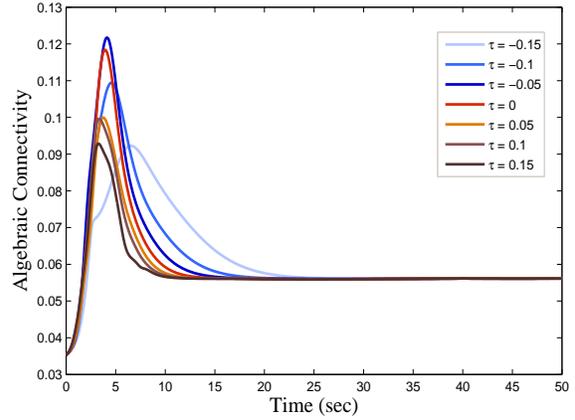}\\
  \caption{Connectivity profiles with varying parameter $\tau$ for initial condition B in Table \ref{tab:initcond}.}
\end{figure}


\section{Conclusion}\label{sec:conclusion}
This paper presents a novel distributed controller for the task of symmetric target-centric formation control and connectivity maintenance. The controller with proper fractional powers on the proportional and derivative difference terms, generates smooth trajectories for a team of UAVs and preserves connectivity. We prove that the proposed controller generates a symmetric formation keeping UAVs connected throughout the formation process. The fractional power controller causes moderate changes in the UAV velocity and step size, avoiding abrupt changes in the time-varying connectivity profile. The formation controller requires comparatively lower range of inputs to accomplish the cooperative task, making it very useful in real applications. In addition, we also presented a fractional power nonlinear observer to estimate the velocity of every UAV at a faster rate than a linear observer.

\bibliographystyle{plain}        

\end{document}